\documentclass[english,12pt]{article}
\usepackage{times}
\usepackage[T1]{fontenc}
\usepackage[latin1]{inputenc}
\usepackage{babel}
\usepackage{lscape}
\usepackage{a4}
\usepackage{pslatex}
\usepackage{graphicx}
\usepackage{amsmath}
\usepackage{verbatim,subfigure}
\usepackage{amssymb,pstricks,enumerate,epsfig}

\begin{document}

%\author{Hindrik Jan Angerman}
%\email[corresponding author: ]{angerman@ics.u-strasbg.fr}
%\affiliation{Institute Charles Sadron (LEA), 6 Rue Boussingault, 67083, Strasbourg, France}
%\author{Albert Johner}
%\affiliation{Institute Charles Sadron, 6 Rue Boussingault, 67083, Strasbourg, France}
%\author{Alexander N. Semenov}
%\affiliation{Institute Charles Sadron, 6 Rue Boussingault, 67083, Strasbourg, France}

\author{Hindrik Jan Angerman$^\star$\thanks{Corresponding author: angerman@ics.u-strasbg.fr;}, Albert Johner$^\dag$, Alexander N. Semenov$^\dag$\\\\
{\em \small $^\star$ Institut Charles Sadron (LEA), 6 rue
  Boussingault, 67083 Strasbourg, France}\\
{\em \small $^\dag$ Institut Charles Sadron, 6 rue
  Boussingault, 67083 Strasbourg, France}
}

\title{Microphase separation in thin block copolymer films:\\
a weak segregation mean-field approach}

\maketitle

\begin{abstract}
In this paper we consider thin films of AB block copolymer melts confined between two parallel plates. The plates are identical and may have a preference for 
one of the monomer types over the other. The system is characterized by four parameters: the Flory-Huggins $\chi-$parameter accounting for the repulsion between
the $A-$monomers and the $B-$monomers, the fraction $f$ of $A-$monomers in the block copolymer molecules, the film thickness $d$, and a parameter $h$ quantifying
the preference of the plates for the monomers of type $A$.
In certain regions of parameter space, the film will be microphase separated. Various structures have
been observed experimentally, each of them characterized by a certain symmetry, orientation, and periodicity. 
We study the system theoretically using the weak segregation approximation to mean field theory.
We restrict our analysis to the region of the parameter space where the interaction strength $\chi$ is close to its spinodal value $\chi_s$, 
the composition $f$ is close to $0.5$, the plate preference $h$ is small, and the film thickness $d$ is close to a small multiple of
the natural periodicity $L_0 = 2 \pi / q_0$ (the first three conditions define the weak segregation regime). 
We will present our results in the form of phase diagrams in which $\Delta \sim | \, (d/n L_0-1) / \epsilon \, |$ is placed along the horizontal axis, and 
$T \sim (\chi-\chi_s)/\epsilon^2$ is placed along the vertical axis, where $\epsilon$ is a measure for the composition asymmetry $(f-0.5)$. 
We present a series of such phase diagrams for increasing values of the rescaled parameter $H = h \epsilon^{-3}$. We find that if $\Delta$ is small, corresponding to films 
whose thickness is commensurate with the bulk periodicity, parallel orientations of the structures are favoured over perpendicular orientations. We also predict that 
on increasing the value of the parameter $H$, the region of stability of the bcc phase shrinks. 
\end{abstract}

%\pacs{61.25.Hq, 61.30.Hn}

\section{Introduction}

Bulk systems of AB block copolymers may undergo a phase transition to a state which is
homogeneous on a macroscopic scale, but which is phase separated on a mesoscopic scale.  
Many experimental \cite{macromolecules_19_2197_1986,macromolecules_27_6922_1994} and theoretical 
\cite{macromolecules_13_1601_1980,macromolecules_26_6617_1993,Journal_of_Chemical_Physics_106_2436_1997,Macromol.Symp._81_253_1994} 
studies of bulk block copolymers have been published, enhancing our understanding of their 
phase behavior. More recently, there has been an increasing interest in thin block copolymer films. 
Such films can be created by spin-coating a block copolymer solution 
on a substrate, and then letting the solvent evaporate. One can create either a free film for which the
upper surface is formed by air while the lower surface is formed by the substrate 
\cite{PRL_62_1852_1989,macromolecules_33_5702_2000,PRL_87_035505_2001,macrocolecules_38_4311_2005}, 
or a confined film for which the upper surface is a solid plate parallel to the substrate \cite{PRL_72_2899_1994}.
There is a qualitative difference between these two cases. If the film is confined, the microstructure might have
to be compressed or stretched in order to fit inside the film.
If the film is free, however, the system can avoid this distortion by 
the formation of terraces \cite{macromolecules_28_7501_1995, PRL_87_035505_2001}, so that locally its thickness is everywhere commensurate with the natural
periodicity of the structure. In this paper we will only consider confined films.

Using X-ray scattering, various microstructures have been observed in copolymer films. The simplest of these is the lamellar structure
\cite{PRL_62_1852_1989,PRL_72_2899_1994,macromolecules_33_5702_2000}.
The lamellae may be oriented either parallel, or perpendicular to the film. 
In \cite{macromolecules_27_6225_1994,Journal_of_Chemical_Physics_106_7781_1997,macromolecules_30_3097_1997,Macrocol_Theory_Simul_7_249_1998,
Journal_of_Chemical_Physics_111_5241_1999,European_physical_journal_E_5_605_2001} 
the relative stability of these two lamellar structures was investigated theoretically for the case of confined films. 
Also more complex structures have been reported experimentally \cite{macrocolecules_38_4311_2005}, such as
cylindrical structures, hexagonally perforated lamellae, and the gyroid structure. The cylinders of the cylindrical structure
can be oriented either parallel, or perpendicular to the film. The relative stability of these structures has been investigated 
theoretically in \cite{Journal_of_Chemical_Physics_112_2452_2000,Journal_of_Chemical_Physics_118_905_2003,physical_review_E_63_061809_2001}.

Apart from experimental and theoretical research, also some Monte Carlo studies have appeared 
\cite{Journal_of_Chemical_Physics_111_5241_1999,Journal_of_Chemical_Physics_118_905_2003}. In \cite{Journal_of_Chemical_Physics_111_5241_1999}
the phase behavior of symmetric AB-diblock copolymers confined in a thin film was investigated using both Monte Carlo simulations, and self-consistent
field theory. The film boundaries were assumed to have a preference for the minority component. For thick films and small AB-incompatibility the
authors found a lamellar structure in which the lamellae are oriented parallel to the film. For thin films and larger incompatibility, the authors
find transitions between parallel and perpendicular orientations of the lamellae. In \cite{Journal_of_Chemical_Physics_118_905_2003} an $A_{8} B_{48} A_{8}$ 
triblock copolymer trapped between two plates having a preference for the end blocks was studied using Monte Carlo techniques. It was found that
in very thin films, a cylindrical structure with the cylinders perpendicular to the film arises, while in thicker films the perpendicular
cylinders are replaced by either parallel cylinders, or parallel lamellae, of perforated lamellae. 

For experimentalists it would be useful to know in advance in which
part of the multi-dimensional phase space one can expect to find interesting microstructures. This kind of information can be
obtained using theory or simulation. We can distinguish two types of theories. On the one hand we have theories which require 
heavy numerical calculations, like the self-consistent field theory. For a given point in phase space they will
give an accurate prediction for the structure to be expected, but since these methods are rather time consuming, they are not
suitable for obtaining an overview over the whole of phase space. On the other hand there exist the more simple (analytical) 
theories like the strong segregation theory, and the weak segregation theory. Due to the approximations made, their predictions 
are less accurate, but they have the advantage that they can provide physical insight, and that the whole phase diagram can be 
mapped out rather quickly, providing a global overview. The rough results of such theories can then be used as a guide for more 
time consuming numerical methods or simulations, or as a guide for experimental work.

\section{Equivalence between a confined film and a bulk system}
\label{Equivalence_between_a_confined_film_and_a_bulk_system}

We consider a thin film of molten $AB-$diblock copolymer trapped between two parallel imprenetrable plates.
The copolymers are assumed to be flexible, in the sense that the persistence length is much smaller than the
block length. 
Apart from the chain connectivity, the excluded volume of the monomers, and the excluded
volume of the plates, which represent strong interactions, two weak interactions are present in the system. The first
is a short-range interaction between the different monomer types, and is taken into account by means of the Flory-Huggins 
$\chi-$parameter. If the $\chi-$parameter is positive, there is a tendency for the $A-$monomers to separate from the
$B-$monomers.
The second is a short-range interaction between the plates and the monomers.
The range of this interaction may be
either comparable with, or shorter, or longer than the size of a monomer unit, but it must be short compared to the radius of gyration 
of the blocks. 
 If the interaction strength between the 
$A-$monomers and the plates is different from the interaction strength between the $B-$monomers and the plates, there 
will be an accumulation of one of the monomer types near the plates. The presence of these two weak interactions may lead 
to a non-homogeneous spatial distribution of the two monomer types. Our aim is to find the time-averaged shape of the equilibrium 
composition profile in terms of the parameters of the system. Having this goal in mind, it is convenient to switch from 
a microscopic description in terms of the positions of all individual monomers, to a mesoscopic description in terms 
of the coarse grained composition profile $\psi$ defined by

\begin{eqnarray}
\psi (x,y,z) &=& \rho_A (x,y,z) \;-\; f
\end{eqnarray}

where $\rho_A$ is the coarse-grained $A-$monomer concentration, and $f$ is the system averaged $A-$monomers concentration
(we assume that the volume of a monomer of type $A$ equals that of a monomer of type $B$). 
The units have been chosen such that the total density of the system equals unity, from which it follows that $0<f<1$.
Note that there are many microscopic states 
corresponding to a given coarse grained state $\psi$. We will analyze the system by calculating its mean-field free energy $F$. 
We make the assumption that this free energy can be obtained by minimizing a functional $F[\psi]$ having the form

\begin{eqnarray}
\label{additivity}
F[\psi] &=& -S[\psi] \;+\; \Delta F[\psi]
\end{eqnarray}

The functional $F[\psi]$ can be considered as a generalization of the Landau free energy known from the study of ferromagnetic systems. 
All effects concerning the weak interactions (monomer-monomer, monomer-plate) are present in the second contribution $\Delta F$,
which vanishes if these interactions are absent. The first term $S[\psi]$ is the entropy of the coarse grained state $\psi$; that is,  
the logarithm of the sum of all \emph{a priori} probabilities of the microscopic states which are compatible with this coarse grained
state. It takes into account the effects of the strong interactions (excluded volume, chain connectivity), and it corresponds 
to the Landau free energy of a film in which the $A-$monomers are physically indistinguishable from the $B-$monomers. 
It is possible to write down an expression for $S$ in the form of a series expansion in $\psi$. The coefficients of this expansion 
can be related to the single-chain correlation functions of the system in which the weak interactions are absent. We will assume that
the third and higher order correlation functions can be built from the second order ones $G(i, \vec R_1; j, \vec R_2)$ 
(see \cite{macromolecules_13_1601_1980}). In principle, these correlation functions could be calculated by summing the \emph{a priori} 
probabilities of all microscopic states in which, of a given chain, monomer $i$ is at position $\vec R_1$, and monomer $j$ is at position 
$\vec R_2$. Within mean-field theory it can be proven that the correlation functions thus obtained satisfy reflecting boundary conditions
at the plates (see \cite{colloid_interface_sciences_90_86_1982}, where this was established for the first time). For a half-infinite system 
occupying the part of space satisfying $z > 0$, the second order correlation function can be written as

\begin{eqnarray}
\label{reflection_principle}
G(i, \vec R_1; j, \vec R_2) &=& G_{bulk}(i,\; \vec R_1; j,\; \vec R_2) \; + \; G_{bulk}(i,\; \vec R_1; j,\; S \vec R_2) 
\end{eqnarray}

where $S$ represents a reflection in the plane $z = 0$. If equation (\ref{reflection_principle}) is generalized to a film, the right hand side 
will contain an infinite number of terms, obtained from repeated reflection of $\vec R_2$ in the plates. We emphasize
that we do not assume that these reflecting boundary conditions still hold in a system for which the plates have a preference for one 
of the monomer types over the other. We only assume that the Landau free energy of the general system can be written as a sum of
two parts, one being the Landau free energy of the system without weak interactions, and that reflecting boundary
conditions may be assumed in the calculations concerning the latter. 

We introduce a coordinate system for which the $z-$axis is perpendicular to the film, and for which the origin is chosen such that
the film is in between $z=0$ and $z=d$. 
The profile $\psi (x,y,z)$ is only defined within the film; that is, for $0<=z<=d$. We now construct the profile 
$\Psi (x,y,z)$, which is defined in the whole of space, by repeated reflection of the film profile in the plates. 
It can be shown that under the assumptions enunciated above, the free energy density of a film 
with profile $\psi (x,y,z)$ is the same as the free energy density of a bulk system with profile $\Psi (x,y,z)$. 
Since much is already known about the calculation of the free energy density of bulk systems, it is convenient to 
switch from the film to the auxiliary bulk system. The profile of the auxiliary bulk system is invariant under 
translation over a distance $2d$ into the z-direction, and mirror symmetric in the plane $z=0$. Note that it 
follows from this that the profile is mirror symmetric in any plane described by $z = k d$ for integer $k$. 
The symmetries imply that we can express the composition profile as a Fourier series of the following general form:

\begin{eqnarray}
\label{general_form_for_profile} 
\Psi (x,y,z) = \Sigma_{k=1}^{\infty} \hspace{0.1cm} A_k(x,y) \hspace{0.1cm} \cos \frac{k \pi z}{d}
\end{eqnarray}

For even values of $k$, the cosine appearing in equation (\ref{general_form_for_profile}) is symmetric with respect to the midplane of the film, 
while for odd values of $k$, it is antisymmetric. Since the plates are identical, only for even values of $k$ does the corresponding composition 
wave couple to the plates. Consider for given vector $\vec q$ and phase $\phi$ the following combination of plane waves:

\begin{eqnarray}
\label{symmetric_profile}
\Psi_{\vec q, \phi}(x,\,y,\,z) &=& e^{ \, i \phi} \, e^{i \, (q_x x \, + \, q_y y \, + \, q_z z)} \, + \, e^{- \, i \phi} \, 
e^{ - \, i \, ( q_x x \, + \, q_y y \, + \, q_z z)}\; + \\ \nonumber \\
&& e^{- \, i \phi} \, e^{ i \, (- \, q_x x \, - \, q_y y \, + \, q_z z)} \; + \; e^{\, i \phi} \, e^{- \, i \, (- \, q_x x \, - \, q_y y \, + \, q_z z)} 
\nonumber \\ \nonumber \\
&=& 2 \, \cos \, (q_x x \, + \, q_y y \, + \, \phi ) \, \cos ( \, q_z z \, ) \nonumber
\end{eqnarray}

with

\begin{eqnarray}
\label{quantization}
q_z &=& \frac{k \pi}{d}
\end{eqnarray}

We see that any real linear combination of profiles of the form given by equations (\ref{symmetric_profile}) and (\ref{quantization}) has the special form given by
equation (\ref{general_form_for_profile}), and thus possesses the symmetries of an auxiliary bulk system. In fact, the converse is also true: any bulk profile
satisfying the above-mentioned symmetry conditions can be written as a linear combination of functions of the form given by equation (\ref{symmetric_profile}):

\begin{eqnarray}
\label{general_form_for_profile2}
\Psi(x,y,z) &=& \sum_{\vec q} \; A_{\vec q} \; \Psi_{\vec q, \phi_{\vec q}} \, (x,\,y,\,z)
\end{eqnarray}

where the summation over $\vec q$ is restricted to some appropriate quadrant of Fourier space (see equation (\ref{symmetric_profile})),
and the amplitudes $A_{\vec q}$ are real. 
It is easier to express the symmetry requirements in terms of the Fourier transform of the composition profile, which is defined by

\begin{eqnarray}
\Psi_{\vec q} = \int d^3 r \; \Psi(\vec r) \; e^{i \, \vec q \cdot \vec r}
\end{eqnarray}

If $\Psi(\vec r)$ is periodic, then its Fourier transform consists of a discrete collection of $\delta-$peaks, each of which has an amplitude
proportional to the volume of the system. 
Using equation (\ref{symmetric_profile}), we see that in order for the profile to be real and the symmetry conditions to be satisfied, the following 
set of three conditions on the Fourier transform of the composition profile is sufficient (it is also necessary, but we will not prove this)

\begin{eqnarray}
\label{symmetry_conditions}
\Psi \,(-\vec q\,)  &=&  \Psi^{*}\,(\vec q\,)
\\ \nonumber \\ 
\Psi \,(\,-\vec q_{_{//}},\, q_z\,)  &=&  \Psi^{*}\,(\, \vec q_{_{//}},\, q_z\,)
\nonumber \\ \nonumber \\
\Psi \,(\, \vec q_{_{//}},\, q_z\,)  &=&  0 \hspace{4 mm} \text{if not} \hspace{4 mm} q_z = \frac{k \pi}{d} \nonumber
\end{eqnarray}

where $k$ is an integer, the star denotes a complex conjugate, and $\vec q_{_{//}}=(q_x, q_y)$ denotes a vector parallel to the film. 
The first condition expresses the fact that the composition profile must be real-valued, while the second and the third express the symmetry properties.
Note that it follows from the second condition that $\Psi(\vec q)$ is real for any vector which is perpendicular to the film, which means 
that sines are not allowed; only cosines are. For the set of $\vec q-$vectors which attain a non-zero amplitude the second condition implies 
that whenever $(\vec q_{_{//}}, q_z)$ is in the set, then $(-\vec q_{_{//}}, q_z)$ is also in the set, and both have the same amplitude, but 
opposite phases. We stress that the symmetry conditions expressed by equation (\ref{symmetry_conditions}) are imposed on the profile $\Psi$ 
of the auxiliary bulk system, not on the profile $\psi$ of the film, which is the restriction of $\Psi$ to the film. The film profile is not 
constrained by symmetry requirements, though it is constrained by the requirements that it must integrate to zero, and that its value must 
lie between $(-f)$ and $(1-f)$. 

\section{Weak segregation regime}

We will study the film in the weak segregation regime, which is defined by the requirements that the $AB-$interaction strength $\chi$ is close
to its critical value $\chi_{c}$, the $A-$monomer fraction $f$ is close to $1/2$, and the preference of the plates for one of the monomer types 
over the other is only slight. The precise meaning of the words ``close'' and ``slight'' will be given further on. We will also assume
that the film thickness is close to a multiple $n L_0$ of the natural periodicity $L_0=2 \pi / q_0$, where $q_0$ denotes the position of the 
minimum of the second order vertex \cite{macromolecules_13_1601_1980}. In our calculations we will assume that
the structure is periodic with period $d/n$ in the direction perpendicular to the film, by which we mean that if the film is divided into $n$
slices of thickness $d/n$, then each slice is identical as far as the internal structure is concerned. 
This assumption of homogeneity is only justified if the film thickness $d$
is much smaller than the bulk correlation length $\xi$, which depends on the system parameters via \cite{macromolecules_20_2535_1987}

\begin{eqnarray}
\xi &\propto& \frac{R_{g}}{N|\chi - \chi_s|^{1/2}}
\end{eqnarray}

where $R_{g}$ is the radius of gyration of an unperturbed chain in the melt. 
If, on the other hand, the film thickness is much larger than $\xi$, then in the middle of the film the structure will be the equilibrium bulk structure,
which may be different from the structure near the plates. Nevertheless, for any given value of the film thickness $d$, no matter how large, the condition $\xi >> d$ 
will be fulfilled once we are close enough to the critical point. Therefore, in the weak segregation regime our assumption concerning the homogeneity of the
structure in the perpendicular direction does not impose any restriction on the film thickness, and the integer $n$ may take on any value.
For large $n$ we are in the regime $L_0 << d << \xi$. Due to the first inequality, incommensurability effects are negligible.
Due to the second inequality, the free energy contribution coming from the surface preference is spread evenly over the entire film, and since the film is
thick while the surface preference is small, the effects of the surface preference will be negligible as well. We conclude that if $n$ is large, neither the
incommensurability, nor the surface preference plays an appreciable role, so that the structure inside the film will concide with the bulk structure.
For this reason we need only to consider small values of $n$. We stress that for given film thickness
$d >> L_0$ the reasoning given above only applies once we are close enough to the critical point, where $\xi >> d$. If, on the other hand, $\xi \lesssim d$,
the influence of the surface preference penetrates only over a certain distance into the film, and the structure near the midplane may be different from
the structure near the surfaces. It would be interesting to study how one structure slowly transforms into another; see also \cite{macromolecules_27_4986_1994}, 
where this problem was addressed in the limit $d \rightarrow \infty$.

\section{Free energy}
\label{section2}

The next step is to write down the free energy of the auxiliary bulk system in terms of its composition profile $\Psi$. Since we are in the weak
segregation regime, the free energy may be expanded in powers of the composition profile. It can be shown that in the generic case, 
the terms till the fourth order have the same order of magnitude, while the higher order terms are negligible. This leads to the following approximation to 
the free energy of the auxiliary bulk system (we did not yet include a term accounting for the preference of the plates for one of the monomer types):

\begin{eqnarray}
\label{general_free_energy}
F &=& \frac{1}{2!V} \sum \left ( \Gamma^{(2)}_Q \, - \, 2 \chi\right ) \, \Psi_Q \, \Psi_{-Q} \; + 
\; \frac{1}{3!V^2} \sum \; \Gamma^{(3)}_{Q_1, Q_2, Q_3} \, \Psi_{Q_1} \, \Psi_{Q_2} \, \Psi_{Q_3} \; + 
\nonumber \\ \nonumber \\
&& + \; \frac{1}{4!V^3} \sum \; \Gamma^{(4)}_{Q_1, Q_2, Q_3, Q_4} \, \Psi_{Q_1} \, \Psi_{Q_2} \, \Psi_{Q_3} \, \Psi_{Q_4} 
\end{eqnarray}

The interaction between A-monomers and B-monomers has been taken into account in the usual way by means of the Flory-Huggins $\chi-$parameter. 
If the composition profile is periodic, the Fourier transform $\Psi_Q$ is only non-zero on a discrete lattice, called the reciprocal lattice. 
In the weak segregation regime, the spacing of this lattice is given by the position $q_0$ of the minimum of the second order vertex function 
$\Gamma^{(2)}$. The lattice vectors having length $q_0$, forming what is called the first harmonics sphere, attain a much larger amplitude than 
the vectors of the higher harmonics spheres. For this reason the latter are usually neglected. In this so-called first harmonics approximation 
all vectors appearing in the summations of equation (\ref{general_free_energy}) have the same length $q_0$. Strictly speaking, this can only be 
arranged if the film thickness is a multiple of half the natural periodicity (and even in that case not all structures
can be made to fit; see the next section). For other values of the film thickness, the Fourier vectors which are perpendicular to the film (which will
always be present if the plates have a preference for one of the monomer types over the other) will have 
to be compressed or extended, giving them a length different from $q_0$. However, since we will only consider values for the film thickness which are 
close to a multiple of the natural periodicity, the length of these vectors will still be close to $q_0$. The fact that all vectors appearing in the summations of
equation (\ref{general_free_energy}) have more or less the same length can be used to simplify this equation. The third and fourth order vertices
$\Gamma^{(3)}$  and $\Gamma^{(4)}$ depend only weakly on the angles between their arguments, and may be approximated by angle-independent constants, 
as follows:

\begin{eqnarray}
\label{constant vertices}
\Gamma^{(3)}_{Q_1, Q_2, Q_3} & \rightarrow & \mu \nonumber \\ \nonumber \\
\Gamma^{(4)}_{Q_1, Q_2, Q_3, Q_4} &  \rightarrow & \lambda
\end{eqnarray}

The constants $\mu$ and $\lambda$ will depend on the composition $f$. For the second order vertex this approximation would be too crude, because although
the length of all Fourier vectors will be close to $q_0$ in the regime under study, the relative change in the value of the second order vertex is large
if we change the length of the $q-$vector even slightly, because $\Gamma^{(2)}(q)$ is close to zero for $q=q_0$ (where $q$ denotes the length of the vector $\vec q$).
Nevertheless, it is not necessary to take into account the full $q-$dependence of the second order vertex; it will be sufficient to make a quadratic
approximation around its minimum $q=q_0$, as follows:

\begin{eqnarray}
\label{parabolic approximation to second order vertex}
\Gamma^{(2)} (q) - 2 \, \chi & \approx & c (q-q_0)^2 \, - \, \tau
\end{eqnarray}

The parameter $\tau$ represents the distance to the spinodal, and it is a function both of $\chi$, and of $f$.
The free energy density of the film can be obtained from that of the auxiliary bulk system by adding a term $E_{\text{plates}}$ accounting for the interaction 
of the monomers with the plates (see also equation \ref{additivity}). Note that we are talking about the weak interaction related to the preference of the plates
for one of the monomer types over the other, not the excluded volume interaction. Under the assumption that this weak part of the monomer-plate interaction
is pair-wise additive and short-ranged (the range should be much smaller than the radius of gyration of the blocks), the corresponding contribution 
to the free energy coming from one of the plates will be linear in the $A-$monomer fraction near this plate, averaged over its 2-dimensional surface. 
We thus obtain for some constant $h_0$

\begin{eqnarray}
\label{wall_interaction}
E_{\text{plates}} &=& -\; h_0 \; \langle \, \Psi(x,\,y,\,0) \, \rangle_{x,\,y} \; - \; h_0 \; \langle \, \Psi(x,\,y,\,d) \, \rangle_{x,\,y}
\end{eqnarray}

We will not attempt to express $h_0$ in terms of more 
basic quantities, but take it as a phenomenological parameter. Adding the various terms together we obtain the following expression for the free energy:

\begin{eqnarray}
\label{3}
\frac{\lambda F}{V} &=& \frac{1}{2} \hspace{0.1cm} \sum_{Q} \hspace{0.1cm} ( \Gamma^{(2)}_Q - 2 \chi ) \, x_Q \, x_{-Q} \; - \; 
\frac{\epsilon}{6} \hspace{0.1cm} \sum_{\{Q_i\}} \hspace{0.1cm} x_{Q_1} \, x_{Q_2} \, x_{Q_3} \; + \nonumber \\ \nonumber \\ \nonumber \\  
&&\; + \; \frac{1}{24} \hspace{0.1cm} \sum_{\{Q_i\}} \hspace{0.1cm} x_{Q_1} \, x_{Q_2} \, x_{Q_3} \, x_{Q_4} \; - 
\; \frac{h L_0}{d} \hspace{0.1cm} \sum_{Q}{}^{*} \hspace{0.1cm} x_Q
\end{eqnarray}

where $Q_i$ for $i=1,2,3,4$ represents a lattice vector, and the summations are restricted to sets of lattice vectors whose sum equals zero.
The star in the last summation indicates that it is restricted to lattice vectors which are perpendicular to the film, and whose length 
satisfies equation (\ref{quantization}) for even values of $k$. The variables $x_Q$ are rescaled amplitudes defined by (see equation 
(\ref{general_form_for_profile2}) for the (implicit) definition of $A_Q$)

\begin{eqnarray}
\label{7}
A_Q = \frac{x_Q}{\sqrt{\lambda}}
\end{eqnarray}

The constant $h$ is defined in terms of $h_0$ (see equation (\ref{wall_interaction})) by

\begin{eqnarray}
h &=& \frac{h_0 \sqrt{\lambda}}{L_0}
\end{eqnarray}

where $L_0=2 \pi/q_0$ is the natural periodicity. The parameter $h$ is a measure for the preference of the plates for the monomers of type $A$ (if $h>0$) or $B$ (if $h<0$). 
Note that in the last term of equation (\ref{general_free_energy}) the division by $d$ is needed in order to arrive at the free energy density, since
the volume of the film is proportional to $d$. The parameter $\epsilon$ appearing in equation (\ref{3}) has been defined in such a way that it is positive if
the $A-$monomers are in the minority. It is given by (see equation (\ref{constant vertices}))

\begin{eqnarray}
\label{epsilon}
\epsilon &=& - \; \frac{\mu}{\sqrt{\lambda}}
\end{eqnarray}

The parameter $\epsilon$ vanishes for $f=0.5$, and 
close to the critical point its dependence on the composition $f$ and the chain length $N$ may be approximated by

\begin{eqnarray}
\label{epsilon_as_a_function_of_f}
\epsilon &\approx& - \; \frac{K}{\sqrt{N}} \; ( f - \frac{1}{2})
\end{eqnarray}

where $K$ is a constant. 

\section{Microstructures in thin films}

We have remarked in section (\ref{Equivalence_between_a_confined_film_and_a_bulk_system}) that if $h=0$, then in the weak segregation regime the free 
energy density of a film equals the free energy density of the bulk system whose structure is obtained by repeated reflection of the film in the 
interfaces. Suppose that for a given film thickness there exists a film
structure which would, upon repeated reflection in its interfaces, lead to the equilibrium bulk structure. It follows 
that this film structure would minimize the film free energy. Such film structure has to be a slice of the equilibrium bulk structure, but the
opposite is not true: not every slice of the equilibrium bulk structure whose thickness coincides with the film thickness would lead back to the 
equilibrium bulk structure upon repeated reflection in its interfaces.
For a given film thickness $d$ such slices may not even exist, depending on the periodicity $L_0$ and the symmetry of the equilibrium bulk structure. 
There is no general rule involving the periodicity; for instance, the condition $d = k L_0$ for some positive integer $k$, is neither necessary nor sufficient.
The fact that this condition is not sufficient becomes clear if we consider the gyroid structure. 
Its weak segregation Fourier transform contains twelve plane waves pointing into various directions, so that in any direction the gyroid structure 
is either non-periodic (but non-constant), or it has a period which is larger than twice the natural periodicity. It follows that there is no 
symmetric slice whose thickness equals the natural periodicity. The fact that the above-mentioned condition is not necessary becomes clear if we 
consider the lamellar structure. For a given periodicity, any slice which is cut perpendicular to the lamellae would lead back to the same lamellar 
structure upon repeated reflection in the plates, irrespective of the thickness of the slice. 

In order to avoid cumbersome formulations, we will say that a slice of a bulk structure is 'symmetric' if it leads back to the same bulk structure 
upon repeated reflection in its interfaces. Using this terminology, we can summarize the above by saying that if the equilibrium bulk structure allows 
for a symmetric slice having thickness $d$, then the equilibrium film structure for $h=0$ is given by the structure inside the slice. In this form the 
principle is not very powerful, not only because its applicablity is restricted to the case $h=0$, but also because symmetric slices exist only for
a discrete set of film thicknesses. 
However, it does suggest a way to search for the equilibrium film structure in the more general situation. The idea is as follows. First we choose 
a bulk structure, and we try to find a symmetric slice whose thickness is close to the film thickness. We compress or expand the bulk structure in
the direction perpendicular to the slice until the thickness of the slice coincides with that of the film. We may also want to compress the structure in some
direction(s) parallel to the film in order to reset the lengths of the oblique Fourier vectors back to their natural value $q_0$. Finally we calculate the 
free energy of the resulting deformed structure, adding a term to account for the surface preference. We repeat this procedure for various bulk structures, 
and compare their free energies. 

We have to search for symmetric slices of small integer thickness $n$ (from now on we will measure lengths in units of the natural periodicity $L_0=2 \pi / q_0$). 
This search is facilitated by considering the bulk structures in Fourier space. The Fourier transform of a periodic structure is only non-zero on 
a discrete set of Fourier vectors, which is called the reciprocal lattice. In a previous section we have given a
symmetry condition on the reciprocal lattice in order for the associated structure to have a symmetric slice whose lower interface coincides with the 
plane $z=0$. This symmetry condition is that if $(\vec q_{//}, q_z)$ is in the reciprocal lattice, then $(-\vec q_{//}, q_z)$, $(\vec q_{//}, -q_z)$, 
and $(-\vec q_{//}, -q_z)$ must also be in the lattice. In addition, the $z-$component of any wave vector must be a multiple of $\pi/d$. 
Since the film thickness is given by $d=n L_0$, where $L_0=2 \pi / q_0$, the $z-$components must be multiples of $q_0/2n$. 
For instance, if the slice has unit thickness, there only three allowed values for the angle $\alpha$ between the vector and the film, 
namely $\alpha = 0$, $\alpha = \pi / 6$, and $\alpha = \pi / 2$.  

We will search for symmetric slices in the lamellar structure, the hexagonal structure, the bcc structure, 
and the fcc structure, which comes down to finding orientations of their respective reciprocal lattices satisfying the above-defined conditions.
If a symmetric orientation does not have perpendicular vectors, we will have to add them by hand, 
because such vectors will always have a non-zero amplitude if the plates prefer one of the monomer types over the other. 

\begin{figure}[t]
\begin{center}
\includegraphics[width=120mm]{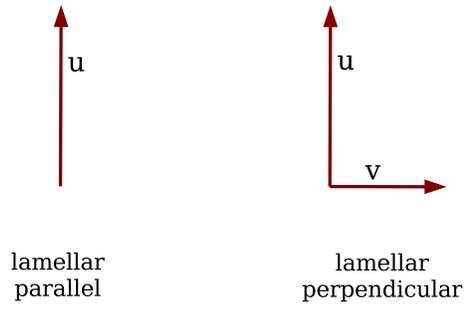}
\caption{Fourier vectors and amplitudes: lamellar structures}
\label{fourier lamellar}
\end{center}
\end{figure}

\begin{figure}[t]
\begin{center}
\includegraphics*[width=120mm]{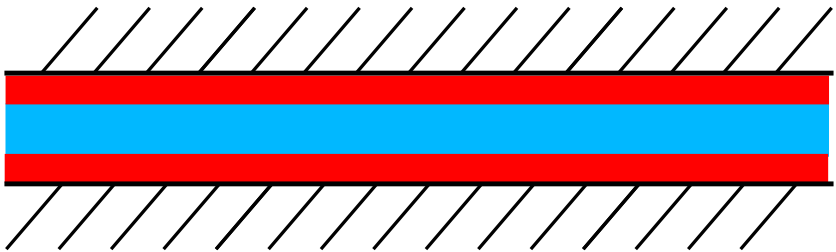}
\caption{Lamellar parallel, side view, n=1}
\label{lamellar parallel side}
\end{center}
\end{figure}

\begin{figure}[t]
\begin{center}
\includegraphics[width=120mm]{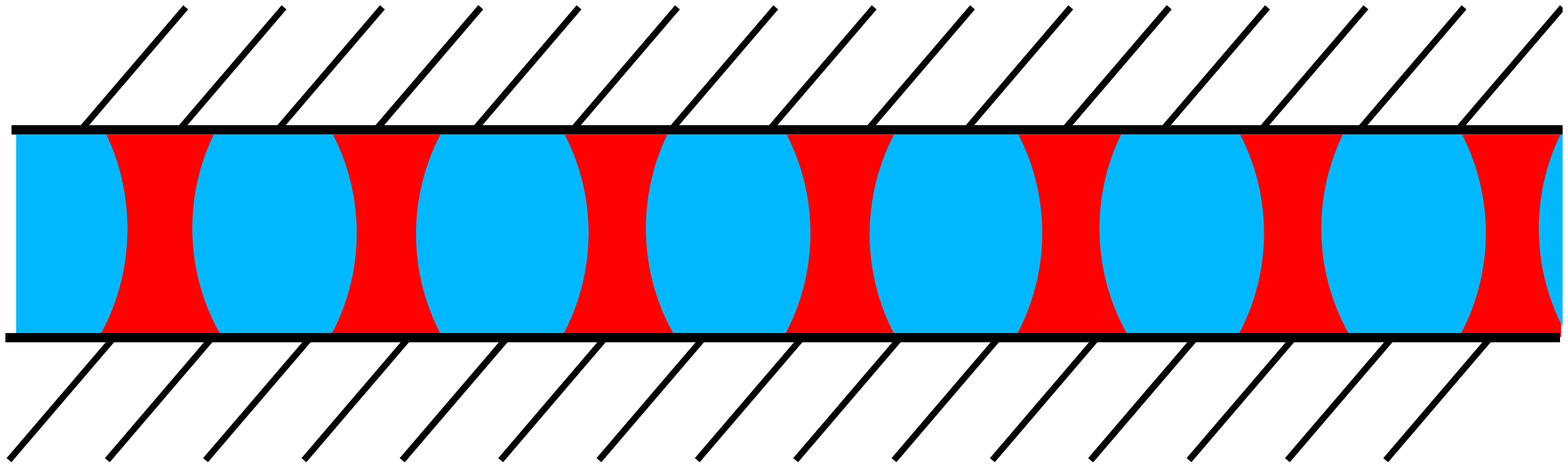}
\caption{Lamellar perpendicular, side view, n=1}
\label{lamellar perpendicular side}
\end{center}
\end{figure}

\begin{figure}[t]
\begin{center}
\includegraphics[width=120mm]{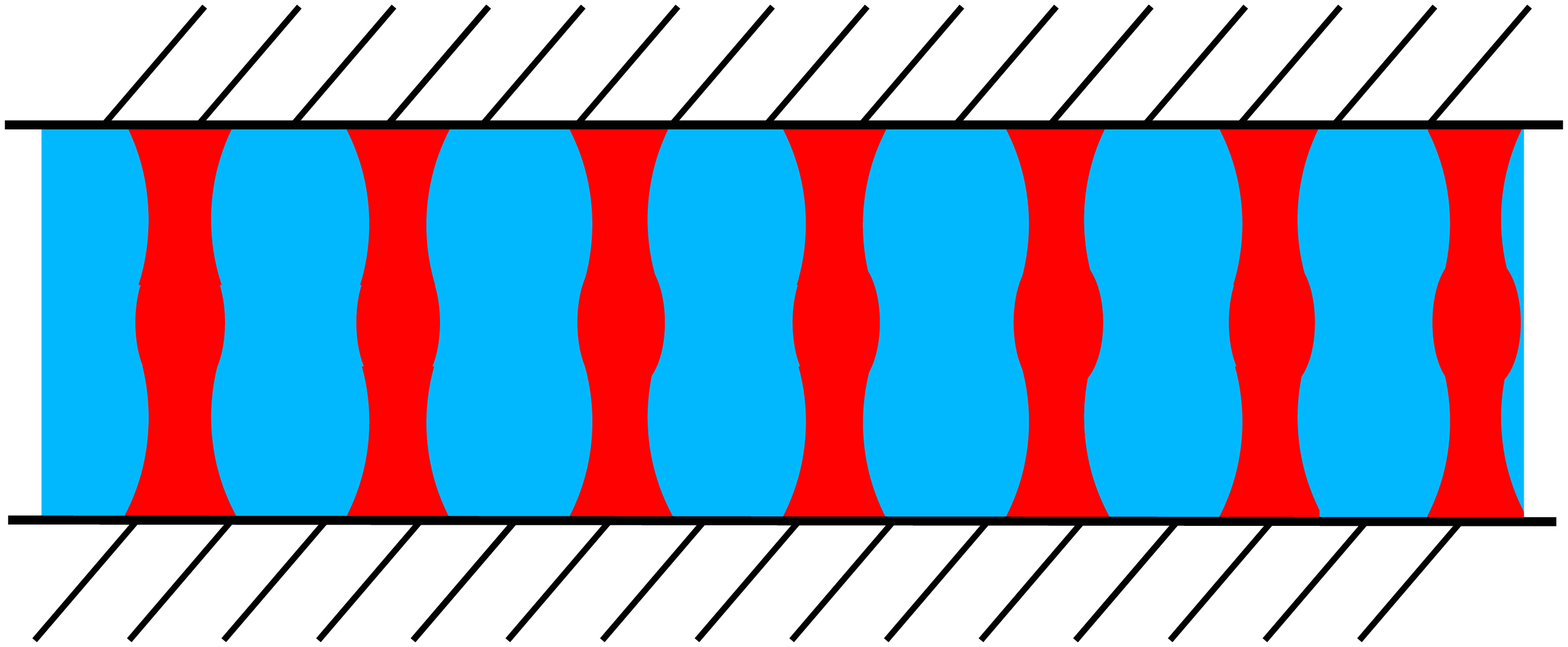}
\caption{Lamellar perpendicular, side view, n=2}
\label{lamellar perpendicular side 2}
\end{center}
\end{figure}

We first consider the lamellar structure. In the first harmonics approximation, its Fourier transform consists of one pair of opposite vectors. 
There are only two ways of satisfying the symmetry conditions formulated above: the vectors should be either perpendicular, or parallel to the film;
see Figure (\ref{fourier lamellar}). Note that in the latter case we have added by hand a set of perpendicular vectors. In real space, 
the first Fourier transform corresponds to lamellae which are oriented parallel to the plates, while the second Fourier transform corresponds to lamellae 
which are oriented perpendicular to the plates. The perpendicular Fourier vectors make it possible that the lamellae have a different thickness near
the plates as compared to near the midplane. Figure (\ref{lamellar parallel side}) shows a cross section of the parallel structure, while Figure
(\ref{lamellar perpendicular side}) shows a cross section of the perpendicular structure, both for the case $n=1$. Figure  
(\ref{lamellar perpendicular side 2}) shows a cross section of the perpendicular structure for the case $n=2$. Note that there is a thickening of
the minority phase near the midplane, which is due to the natural periodicity of the block copolymer.

\begin{figure}[t]
\begin{center}
\includegraphics[width=120mm]{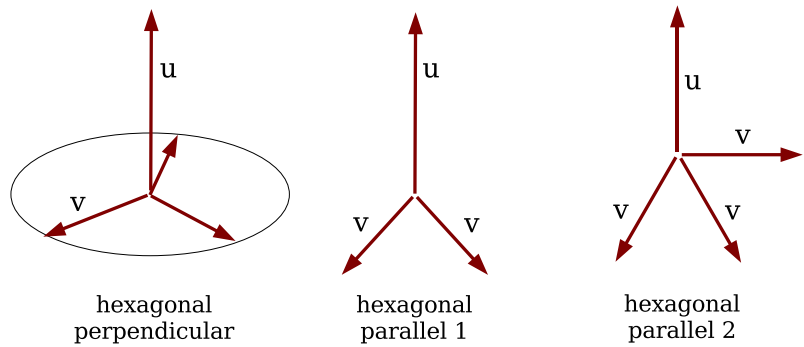}
\caption{Fourier vectors and amplitudes: cylindrical structures}
\label{fourier hexagonal}
\end{center}
\end{figure}

\begin{figure}[t]
\begin{center}
\includegraphics[width=120mm]{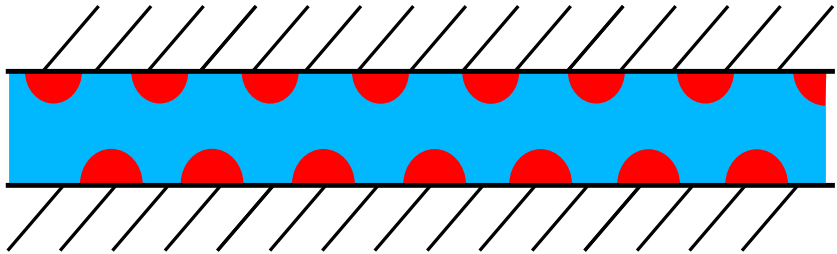}
\caption{Hexagonal parallel, side view, n=1}
\label{hexagonal parallel side}
\end{center}
\end{figure}

\begin{figure}[t]
\begin{center}
\includegraphics[width=120mm]{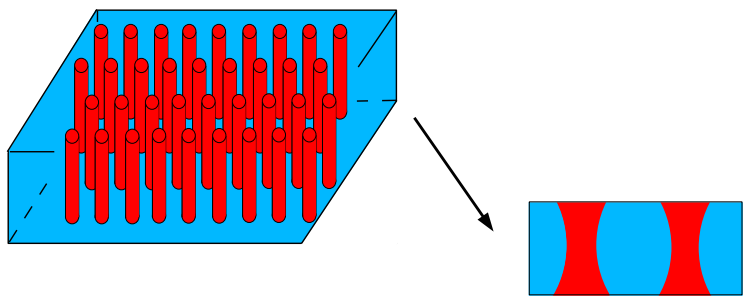}
\caption{Hexagonal perpendicular, n=1}
\label{hexagonal perpendicular}
\end{center}
\end{figure}

Next we consider the structure which consists of hexagonally packed parallel cylinders. In the first harmonics approximation, 
its Fourier transform is a set of six coplanar vectors of length $q_0$, making angles of sixty degrees with each other. There are three ways of orienting 
the vectors with respect to the film such that the symmetry conditions are satisfied, which are illustrated in Figure (\ref{fourier hexagonal}). 
The first possibility is to orient all six vectors parallel to the film, in which case we need an additional set of perpendicular vectors. In real space 
this corresponds to hexagonally packed cylinders oriented perpendicular to the plates. Due to the perpendicular Fourier vectors, the cylinders will have 
a different thickness near the plates as compared to near the midplane (see Figure (\ref{hexagonal perpendicular})). The second possibility is to orient
two of the six vectors vectors perpendicular to the film. Since the $z-$components of the four oblique vectors
are $\pm q_0/2$, this structure will fit without distortion in any film whose thickness is a multiple of the natural periodicity $L_0$. In real space
it corresponds to the cylinders lying parallel to the film, and a cross section for the case $n=1$ is given by Figure (\ref{hexagonal parallel side}). 
The third way to orient the structure is obtained if we rotate the previous (parallel) structure 
over an angle of 30 degrees, so that two Fourier vectors become parallel to the film. We will have to add a set of perpendicular
vectors to account for the plate interaction. The $z-$components of the oblique vectors are equal to $\sqrt{3}/2$ in units of $q_0$, and since this
is an irrational number, the structure will not fit in any film of integer thickness. If $n$ is large enough, it can be made to fit by just a slight
distortion, but the corresponding free energy will always be higher than that of the perpendicular orientation, since both structures have the same
terms in the free energy. It follows that this way of orienting the cylinders is never stable, 
which is in accordance with experimental findings \cite{macrocolecules_38_4311_2005}. 

\begin{figure}[t]
\begin{center}
\includegraphics[width=120mm]{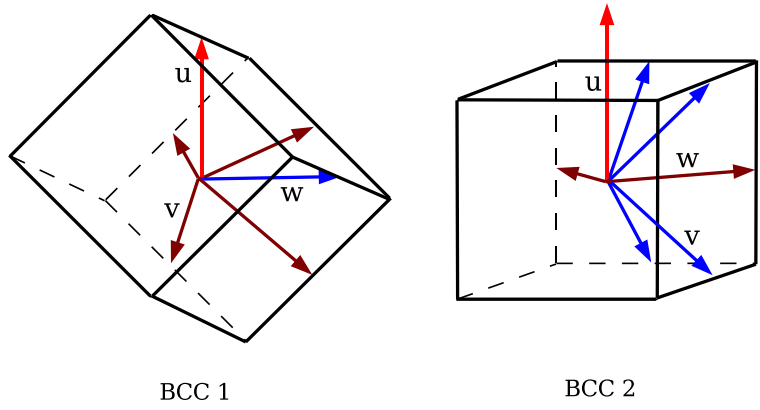}
\caption{Fourier vectors and amplitudes: bcc structure}
\label{fourier bcc}
\end{center}
\end{figure}

\begin{figure}[t]
\begin{center}
\includegraphics[width=120mm]{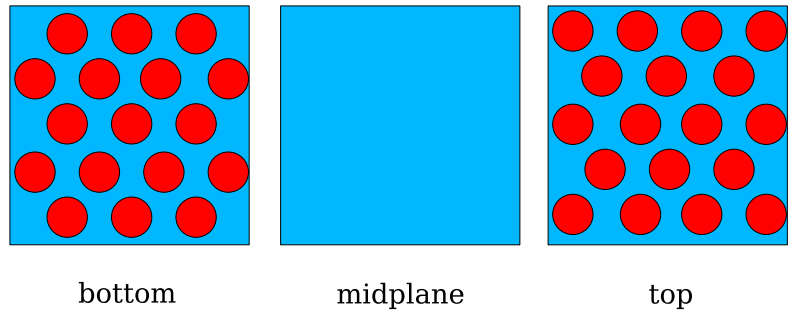}
\caption{BCC top view, n=1}
\label{bcc}
\end{center}
\end{figure}

Now we consider the bcc structure. Apart from trivial rotations around a perpendicular axis, there are only two possiblities to orient the Fourier vectors 
such that the symmetry requirements are satisfied (see Figure (\ref{fourier bcc})). The first possibility is to orient one Fourier vector
perpendicular to the film. All oblique vectors have a $z-$component which is equal to $q_0/2$, and the structure 
fits without distortion in any film of integer thickness. Figure (\ref{bcc}) shows what this structure looks like in real space in a film of unit thickness. 
If the film thickness is slightly increased from integer value, the oblique vectors 
have to be turned slightly towards the horizontal orientation in order for their $z-$component to remain commensurate with the film thickness. 
At the same time, the angle between two oblique vectors whose sum equals one of the horizontal vectors must be increased. Since we ignore the angle dependence 
of the vertices, these distortions do not complicate the calculations. In the second way to orient the Fourier vectors, none of them is perpendicular 
to the film, and perpendicular vectors have to be added in order to account for the plate interaction. The $z-$component of the oblique vectors equals $q_0/\sqrt{2}$, 
and since $\sqrt{2}$ is irrational, the structure will not fit without distortion in any film of integer thickness. This means that for small values of $n$ 
this structure would have a high free energy and may be ignored, but for large values of $n$ it would have to be taken into account, since
the degree of distortion needed to make it fit would be small. In fact, it can be argued that it will become stable in thick films if the plates have a 
preference for the majority component. Nevertheless, for reasons expounded before, we restrict ourselves to small values of $n$, so that we
may neglect the second orientation of the bcc structure. 

There is also a film structure which is derived from the bulk fcc structure. We took it into account in our calculations, but since it is nowhere stable, 
we will refrain from discussing it any further. Finally we considered the gyroid structure. Due to its large number of Fourier vectors pointing into various
directions, it does not fit inside thin films (i.e., $n$ small), and can only be embedded in thick films. Since we restrict ourselves to small values of $n$, 
we do not need to consider the gyroid structure.

\section{Rescaled parameters}

Using the results of the previous section, we can write down expressions for the free energy of the various structures.
These expressions will not only depend on the symmetry of the structure and the amplitudes of the Fourier vectors, but also on the degree of distortion 
(compression or stretching),  which can be quantified by the dimensionless parameter $D$ defined by

\begin{eqnarray}
D &=& \frac{d}{n L_0}
\end{eqnarray}

where the integer $n$ is fixed by the requirement that $n L_0$ is as close as possible to $d$.
The structure is compressed if $D<1$, and stretched if $D>1$. 
Since we consider films whose thickness is close to a multiple of the natural periodicity, we have $D \approx 1$. 
We are now in a position to write down the expressions for the free energy of the various structures
in terms of the Fourier amplitudes defined by Figures (\ref{fourier lamellar}), (\ref{fourier hexagonal}) 
and (\ref{fourier bcc}). The result is

\begin{eqnarray}
\label{free energy in original parameters}
\frac{\lambda}{V}\; F_{lm}  \; &=& \; - \, \tilde{\tau} \; u^2 \; + \; \frac{1}{4} \; u^4 \; - \; \frac{2 h u}{n D} \nonumber \\ \nonumber \\
\frac{\lambda}{V}\; F_{sqr} \; &=& \; - \, \tilde{\tau} \, u^2 \; - \; \tau \, v^2 \; + \; \frac{1}{4} \, u^4 \; + 
\; \frac{1}{4} \, v^4 \; + \; u^2 v^2 \; - \; \frac{2 h u}{n D} \nonumber \\ \nonumber \\
\frac{\lambda}{V}\; F_{hex_{_{//}}} \; &=& \; - \, \tilde{\tau} \, u^2 \; - \; 2 \, \tau \, v^2 \; - \; 2 \, \epsilon \, u \, v^2 \; + 
\; \frac{1}{4} \, u^4 \; + \; \frac{3}{2} \, v^4 
\; + \; 2 u^2 v^2 \; - \; \frac{2 h u}{n D} \nonumber \\ \nonumber \\
\frac{\lambda}{V}\; F_{hex_{_{\frac{|}{\hspace{2mm}}}}} \; &=& \; - \, \tilde{\tau} \, u^2 \; - \; 3 \, \tau \, v^2 \; - 
\; 2 \, \epsilon \, v^3 \; + \; \frac{1}{4} \, u^4 \; + \; \frac{15}{4} \, v^4 
\; + \; 3 u^2 v^2 \; - \; \frac{2 h u}{n D} \nonumber \\ \nonumber \\
\frac{\lambda}{V}\; F_{bcc} \; &=& \; - \, \tilde{\tau} \, u^2 \; - \; \tau \, v^2 \; - \; 4 \, \tau \, w^2 \, -
\; 4 \, \epsilon \, u \, w^2 \; - \; 4 \, \epsilon \, v \, w^2 \; + \; 
\frac{1}{4} \, u^4 \; + \; \frac{1}{4} \, v^4 \nonumber \\ \nonumber \\ && \; + \; 9 \, w^4 
\; + \; u^2 v^2 \; + \; 4 \, u^2 w^2 \; + \; 4 \, v^2 w^2 \; + \; 4 \, u v w^2 \; - \; \frac{2 h u}{n D} 
\end{eqnarray}

The parameter $\tau$ has been defined in equation (\ref{epsilon_as_a_function_of_f})), and is given by

\begin{eqnarray}
\label{tau}
-\tau &=& \Gamma^{(2)} (q_0) - 2 \chi
\end{eqnarray}

while $\tilde\tau$ is defined by

\begin{eqnarray}
\label{tau tilde}
- \, \tilde\tau &=& \Gamma^{(2)} (q_0/D) - 2 \chi
\end{eqnarray}

By making use of the fact that $D$ is close to unity, we can simplify the free energy expressions of equation (\ref{free energy in original parameters}),
reducing the number of parameters from four (namely, $f$, $\chi$, $h$, and $D$), to three. First we relate $\tilde \tau$ to $\tau$
by combining equation (\ref{parabolic approximation to second order vertex}) with equation (\ref{tau tilde}):

\begin{eqnarray}
- \, \tilde\tau &=& c \, q_0^2 \, \left ( \frac{1-D}{D} \right )^2 \, - \, \tau \;\; \sim \;\; c \, q_0^2 \, (D-1)^2 \, - \, \tau
\end{eqnarray}

where we have only kept the leading order term in $(1-D)$, which is justified by our assumption that $D$ is close to unity. In the same approximation, the factor
$D$ may be omitted from the linear term describing the interaction between the A-monomers and the plates; that is, in equation (\ref{free energy in original parameters}) 
we may make the substitution

\begin{eqnarray}
\frac{2 h u}{n D} &\rightarrow& \frac{2 h u}{n}
\end{eqnarray}

As a first step in the reduction of the number of parameters, we replace the original set $(f, \chi, h, D)$ by the equivalent set $(\epsilon, \tau, h, D)$. 
Next we rescale the parameters $\tau$, $h$, and $D-1$ by some appropriate power of $\epsilon$, and arrive thus at the new parameters $\Delta$, $T$, and $H$:

\begin{eqnarray}
\label{rescaled parameters}
\Delta = \sqrt{c} \, q_0 \, \frac{|D-1|}{|\epsilon|} \hspace{2cm}
T = \frac{\tau}{\epsilon^2} \hspace{2cm}
H = \frac{h}{\epsilon^3}
\end{eqnarray}

We also rescale the amplitudes

\begin{eqnarray}
\label{rescaled amplitudes}
u \; \rightarrow \; \epsilon \; u \hspace{1.5cm}
v \; \rightarrow \; \epsilon \; v \hspace{1.5cm}
w \; \rightarrow \; \epsilon \;  w 
\end{eqnarray}

and the free energy

\begin{eqnarray}
\label{rescaled free energy}
F & \rightarrow & \frac{V \epsilon^4 F}{\lambda}
\end{eqnarray}

Substitution of equations (\ref{rescaled parameters}), (\ref{rescaled amplitudes}), and (\ref{rescaled free energy}) into equation (\ref{free energy in original parameters}) 
leads to

\begin{eqnarray}
\label{final expressions}
F_{lam} \; &=& \; ( \, \Delta^2 \, - \, T \, ) \, u^2 \, + \, \frac{1}{4} \, u^4 \, - \, \frac{2 \, H \, u}{n} \nonumber \\ \nonumber \\
F_{sqr} \; &=& \; ( \, \Delta^2 \, - \, T \, ) \, u^2 \, - \, T \, v^2  \,+ \,\frac{1}{4} \, u^4 \, + \, u^2 \,v^2 \,+  \,\frac{1}{4}  
\, v^4 \, - \, \frac{2 \, H \, u}{n} \nonumber \\ \nonumber \\
F_{hex_{_{//}}} \; &=& ( \, \Delta^2 \, - \, T \, ) \, u^2 \, - \, 2 \, T \, v^2  \, - \, 2 \, u \, v^2 \,+ \,\frac{1}{4} \, u^4 \, + 
\, 2 \, u^2 \, v^2  \, + \, \frac{3}{2} \, v^4 \, - \, \frac{2 \, H \, u}{n} \nonumber \\ \nonumber \\
F_{hex_{_{\frac{|}{\hspace{2mm}}}}} \; &=& \, ( \, \Delta^2 \, - \, T \, )  \, u^2 \, - \, 3 \,T \, v^2 \, - \, 2 \, v^3 
\, + \, \frac{1}{4} \, u^4 \, + \, 3 \, u^2 \, v^2 \, + \, \frac{15}{4} \, v^4 \, - \, \frac{2 \, H \, u}{n} \nonumber \\ \nonumber \\
F_{bcc} \; &=& \; ( \, \Delta^2 \, - \, T \, ) \, u^2 \, - \, T \, v^2 \, - \, 4 \, T \, w^2 \, - \, 4 \, u \, w^2 \, - 
\, 4 \, v \, w^2 \, + \, \frac{1}{4} \, u^4 \, + \, \frac{1}{4}  \, v^4 \, + \nonumber \\ \nonumber \\ && \, + \, 9 \,w^4 \, + \, u^2 \, v^2 \, + 
\, 4 \, u^2 \, w^2 \, + \, 4 \, v^2 \, w^2 \, + \, 4 \, u \, v \, w^2 \, - \, \frac{2 \, H \, u}{n} \nonumber \\ 
\end{eqnarray}

From the fact that the (small) parameters $\tau$, $\epsilon$, $h$, and $D-1$ appear only in the combinations $T$, $H$, and $\Delta$, it can be concluded that the 
relevant values for these parameters satisfy the scaling

\begin{eqnarray}
\tau \sim \epsilon^2 \hspace{1.5cm} (D-1) \sim \epsilon \hspace{1.5cm} h \sim \epsilon^3
\end{eqnarray}

We note that although the original parameters $(f-0.5)$, $(\chi-\chi_c)$, $(D-1)$, and $h$ all have to be small in order for the 
theory to be valid, expression (\ref{final expressions}) is valid for any values of the parameters $\Delta$, $T$, and $H$.
This can be understood by realizing that we went from four independent parameters to three. There 
must be a fourth independent parameter in addition to $\Delta$, $T$, and $H$, say $X$, on which the free energy 
depends in a trivial manner, so that the ratio of any two free energies does not depend on $X$. This parameter $X$ is a measure
for the distance to the critical point; therefore, it must be small in the weak segregation regime. The remaining three parameters
describe the position on the three-dimensional hyper-surface defined by the condition that $X$ is constant. Varying these
parameters between zero and infinity only changes the position on this hypersurface, without bringing us further away from the
critical point. It follows that in the phase diagrams to be presented in the next section, where $\Delta$ is placed along the
horizontal axis and $T$ along the vertical axis, any region in the phase diagram is meaningful in the weak segregation regime, 
not just the region near the origin. 

\section{Results and Discussion}

We will present our results in the form of phase diagrams. Along the horizontal axis we put $\Delta \sim |D-1|/|\epsilon|$, and along 
the vertical axis we put $T = \tau / \epsilon^2$. Since there is a third parameter, namely $H = h / \epsilon^3$, we need to calculate a 
series of such phase diagrams, for increasing values of $H$. From the definitions of the parameters $\Delta$, $T$, and $H$ we see that
the phase diagram does not change if we flip simultaneously the signs of $\epsilon$ and $h$. 
This is due to the fact that if we exchange $A$ and $B$ in the copolymer molecules, and at the same time reverse the plate preference,
the equilibrium structure is simply inverted according to $\psi \rightarrow -\psi$, keeping the same symmetry and amplitude. 
Therefore, the phase diagrams for 
$H>0$ describe both the situation where the $A-$monomers are in the minority ($\epsilon > 0$) and the plates prefer the $A-$monomers ($h>0$), 
and the situation where the $B-$monomers are in the minority ($\epsilon < 0$) and the plates prefer the $B-$monomers ($h<0$). 
In order to avoid cumbersome formulations, in our discussions we will always assume that the $A-$monomers are in the minority.

\begin{figure}[t]
\begin{center}
\includegraphics*[width=90mm,angle=270]{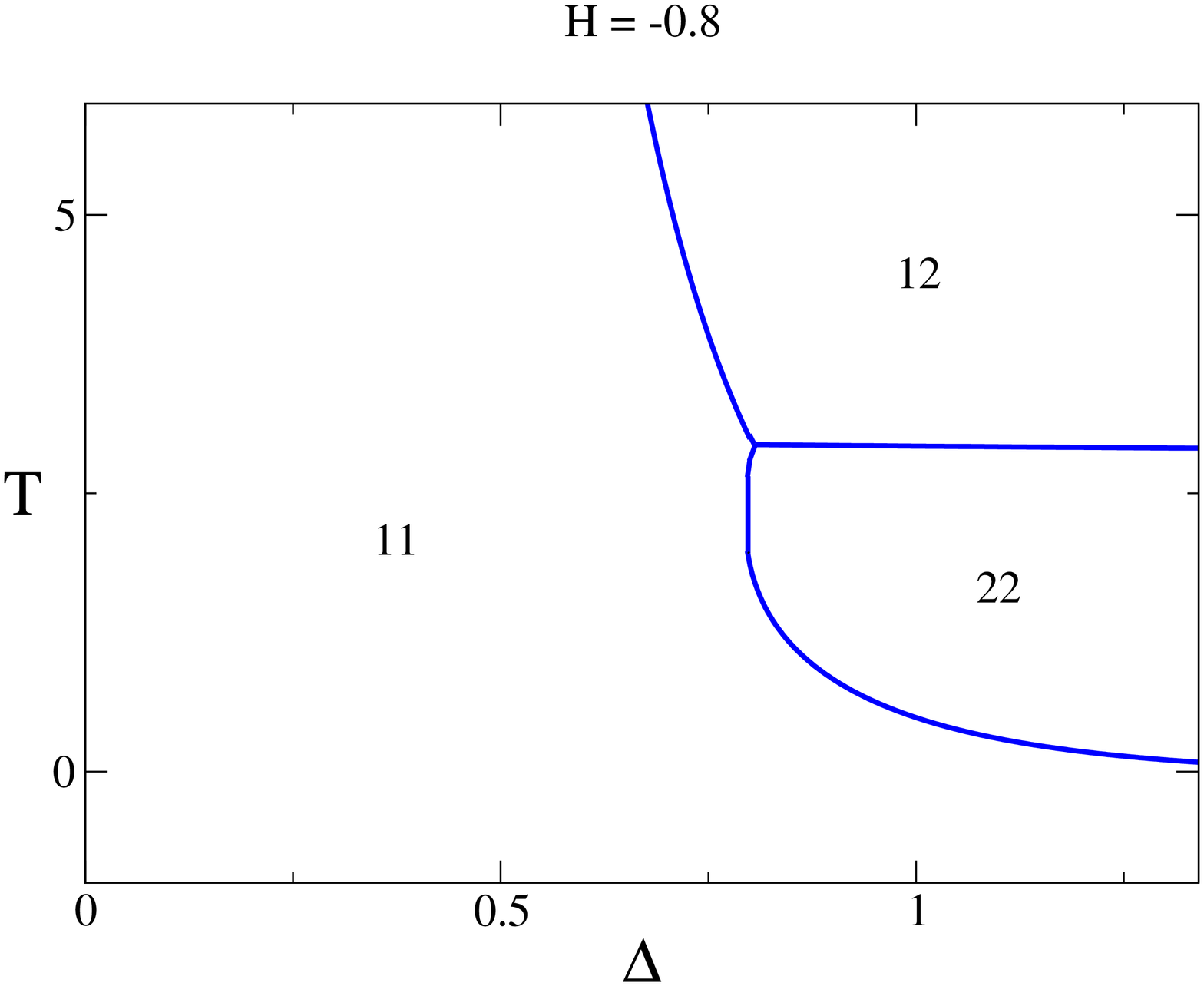}
\caption{Phase diagram for $H = -0.80$.\newline
11 = lamellar parallel,
12 = lamellar perpendicular,\newline
21 = hexagonal parallel,
22 = hexagonal perpendicular,
3  = bcc
}
\label{Hm80}
\end{center}
\end{figure}

\begin{figure}[t]
\begin{center}
\includegraphics*[width=90mm,angle=270]{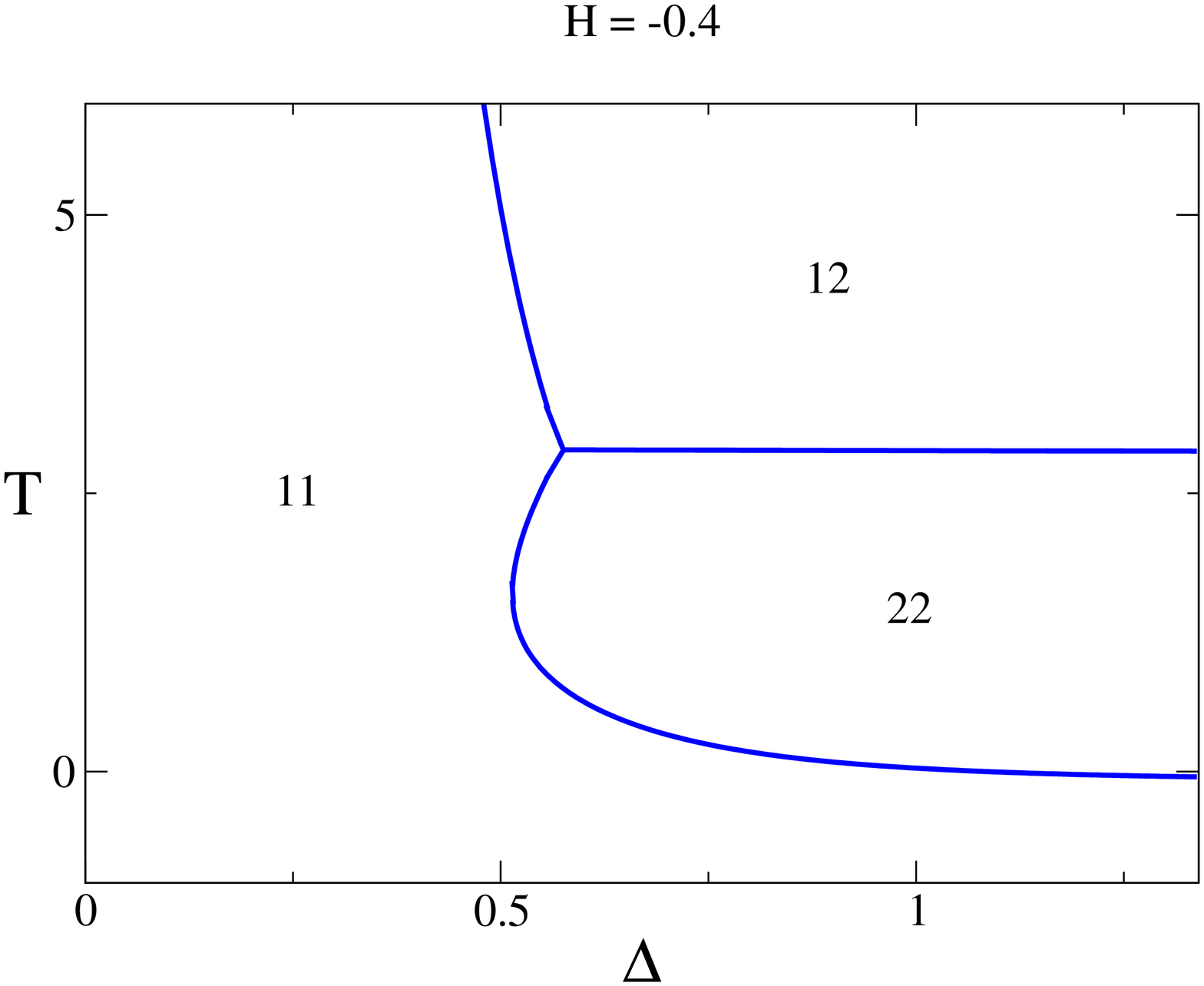}
\caption{Phase diagram for $H = -0.40$.\newline
11 = lamellar parallel,
12 = lamellar perpendicular,\newline
21 = hexagonal parallel,
22 = hexagonal perpendicular,
3  = bcc
}
\label{Hm40}
\end{center}
\end{figure}

\begin{figure}[t]
\begin{center}
\includegraphics*[width=90mm,angle=270]{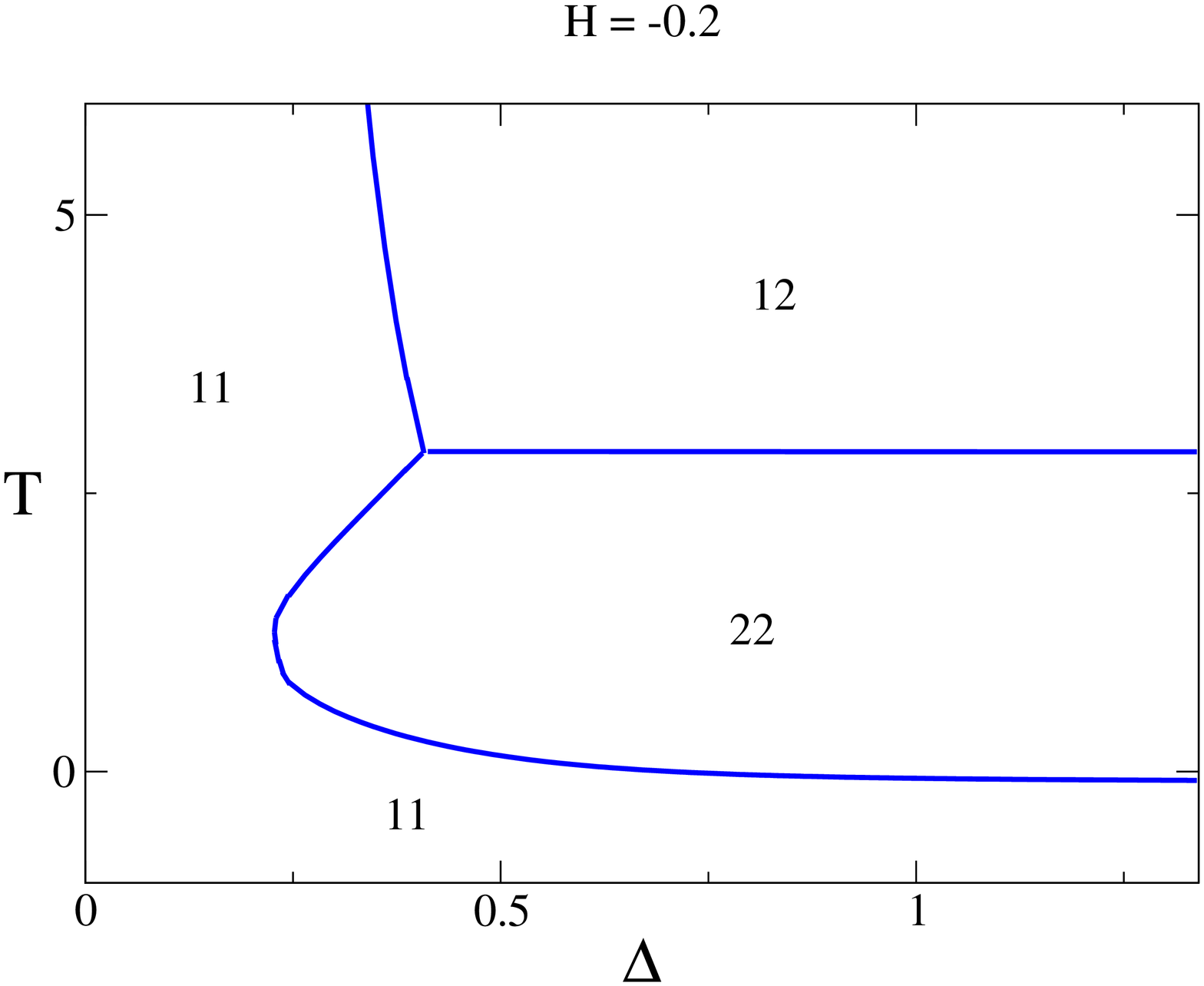}
\caption{Phase diagram for $H = -0.20$.\newline
11 = lamellar parallel,
12 = lamellar perpendicular,\newline
21 = hexagonal parallel,
22 = hexagonal perpendicular,
3  = bcc}
\label{Hm20}
\end{center}
\end{figure}

\begin{figure}[t]
\begin{center}
\includegraphics*[width=90mm,angle=270]{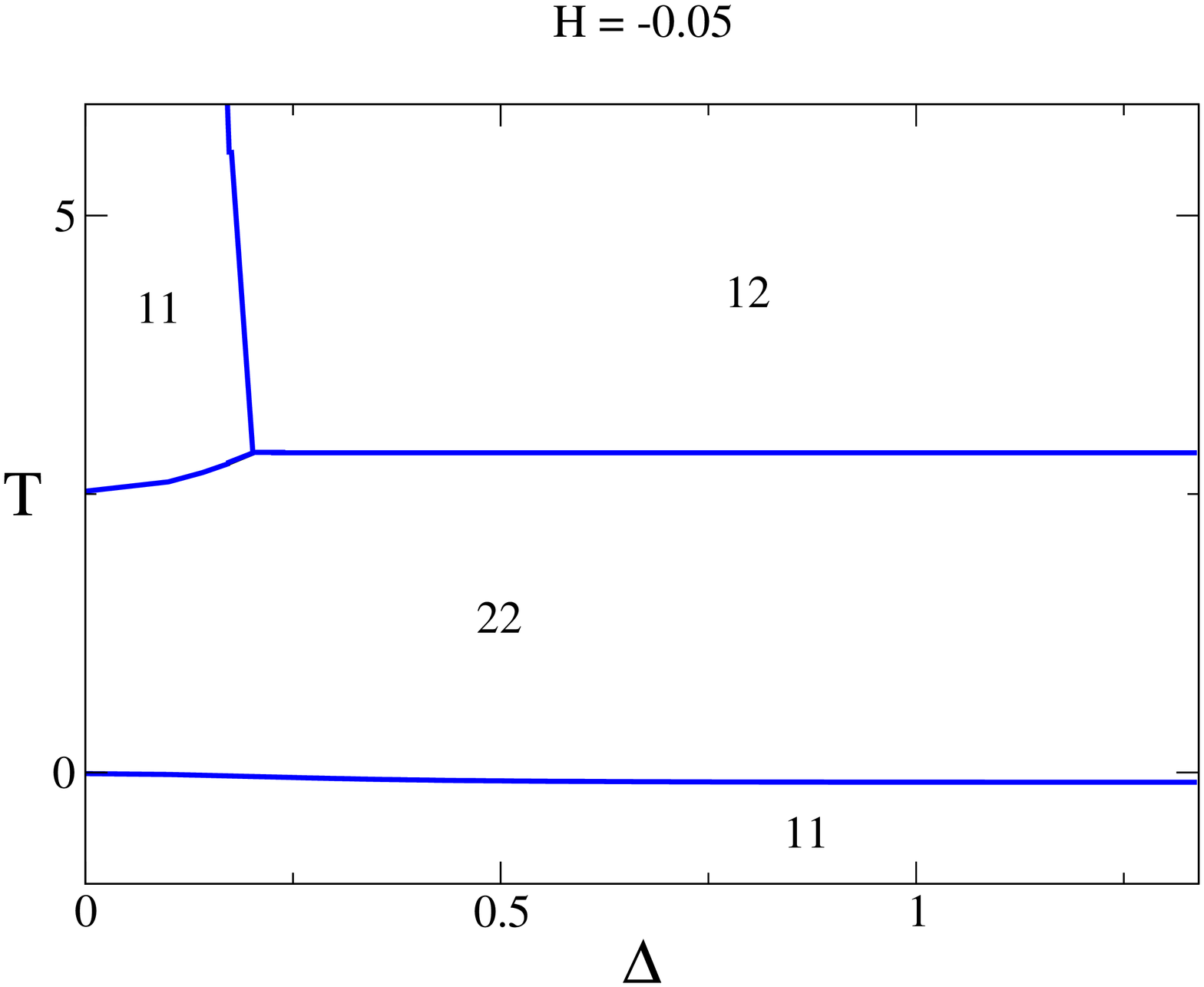}
\caption{Phase diagram for $H = -0.05$.\newline
11 = lamellar parallel,
12 = lamellar perpendicular,\newline
21 = hexagonal parallel,
22 = hexagonal perpendicular,
3  = bcc}
\label{Hm5}
\end{center}
\end{figure}

\begin{figure}[t]
\begin{center}
\includegraphics*[width=90mm,angle=270]{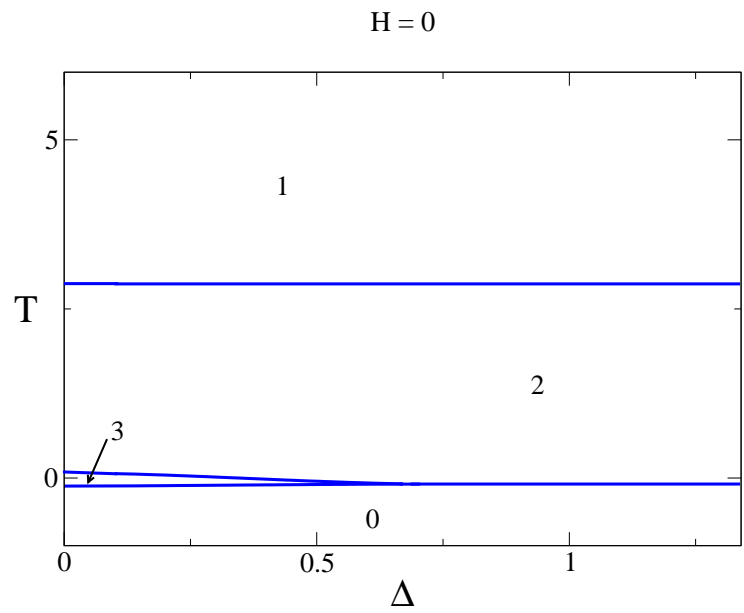}
\caption{Phase diagram for $H = 0$.\newline
0 = homogeneous,
1 = lamellar,
2 = hexagonal,
3 = bcc}
\label{Hp0}
\end{center}
\end{figure}

\begin{figure}[t]
\begin{center}
\includegraphics*[width=90mm,angle=270]{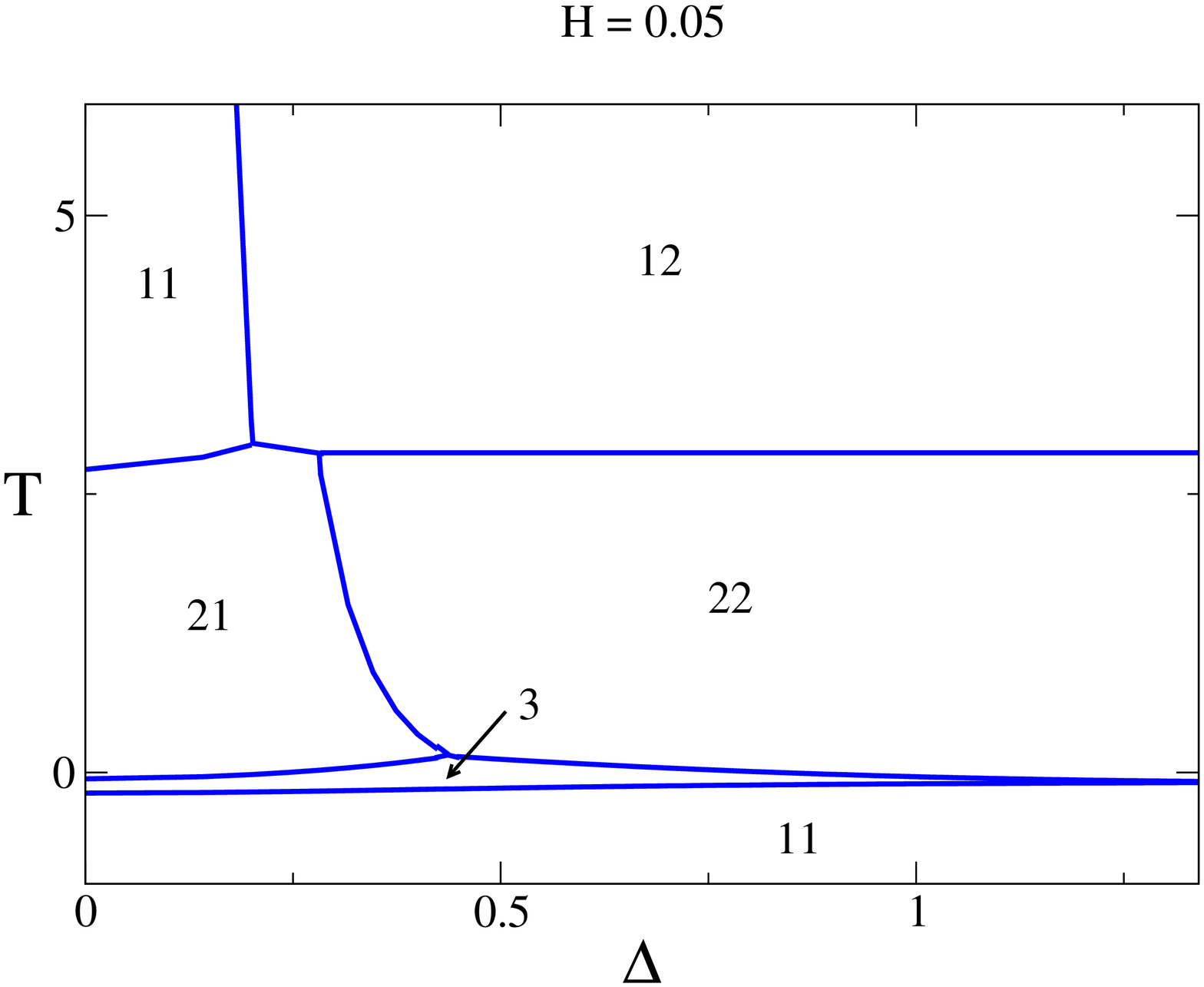}
\caption{Phase diagram for $H = 0.05$.\newline
11 = lamellar parallel,
12 = lamellar perpendicular,\newline
21 = hexagonal parallel,
22 = hexagonal perpendicular,
3  = bcc}
\label{Hp50}
\end{center}
\end{figure}

\begin{figure}[t]
\begin{center}
\includegraphics*[width=90mm,angle=270]{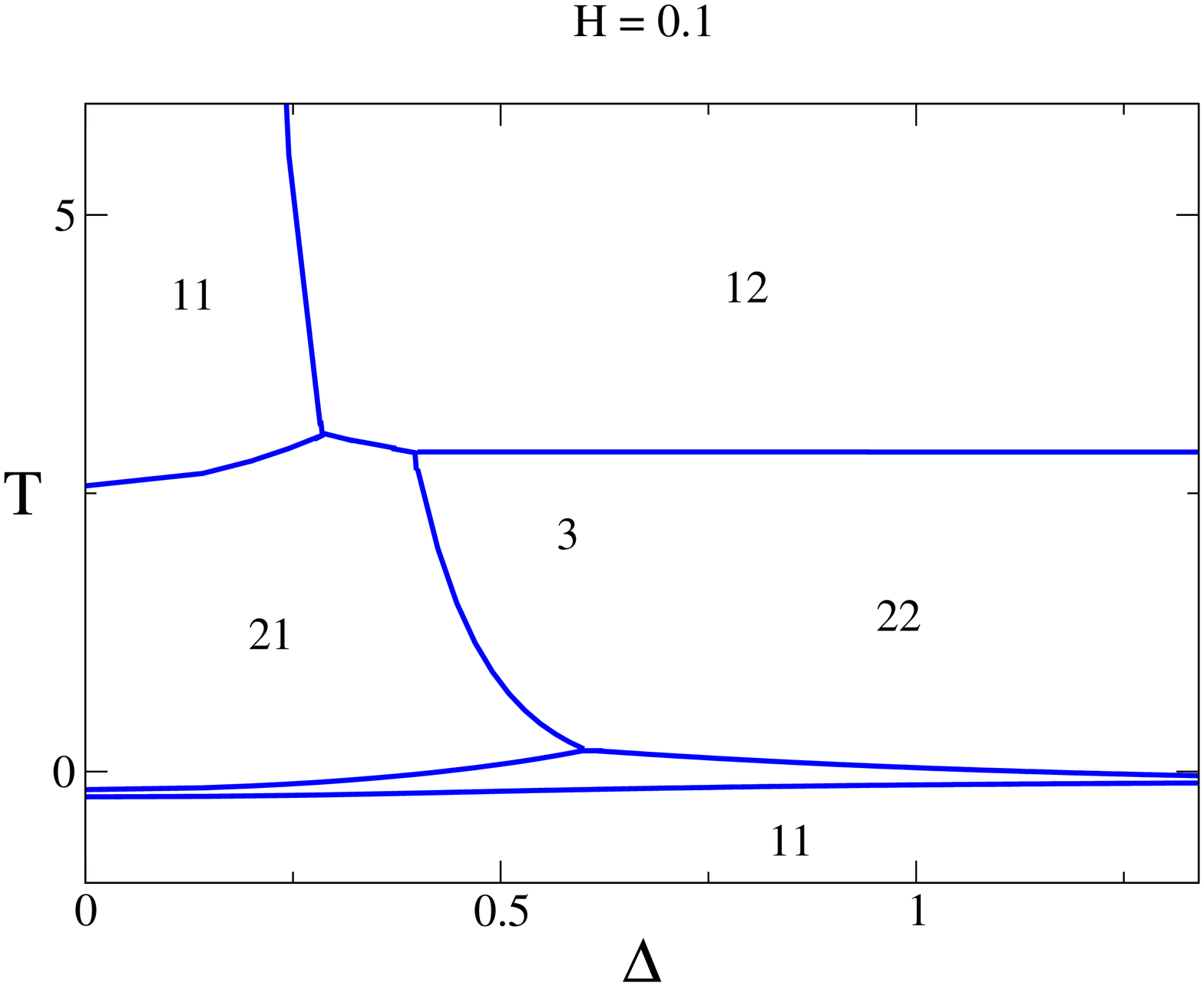}
\caption{Phase diagram for $H = 0.1$.\newline
11 = lamellar parallel,
12 = lamellar perpendicular,\newline
21 = hexagonal parallel,
22 = hexagonal perpendicular,
3  = bcc}
\label{Hp10}
\end{center}
\end{figure}

\begin{figure}[t]
\begin{center}
\includegraphics*[width=90mm,angle=270]{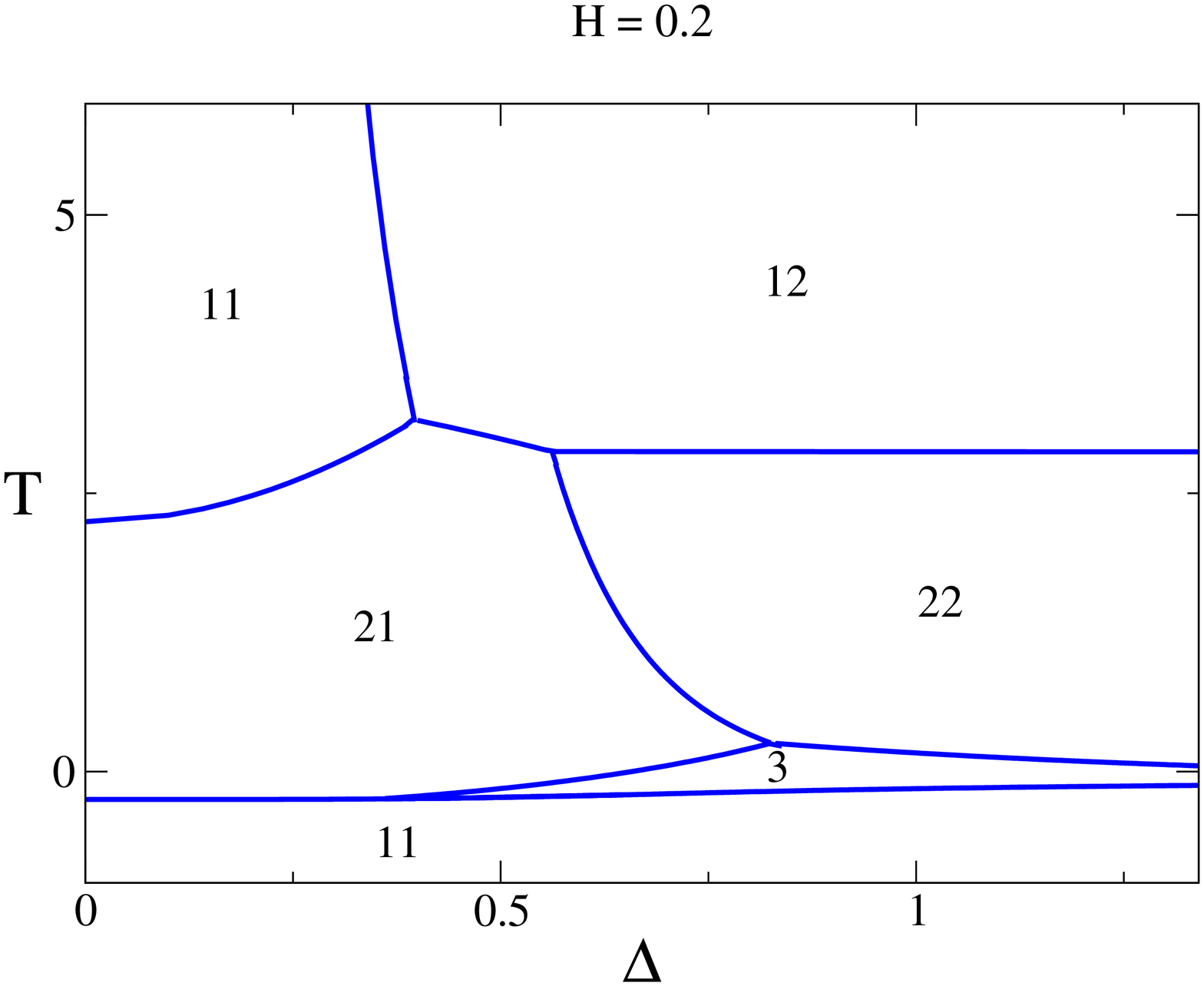}
\caption{Phase diagram for $H = 0.20$.\newline
11 = lamellar parallel,
12 = lamellar perpendicular,\newline
21 = hexagonal parallel,
22 = hexagonal perpendicular,
3  = bcc}
\label{Hp20}
\end{center}
\end{figure}

\begin{figure}[t]
\begin{center}
\includegraphics*[width=90mm,angle=270]{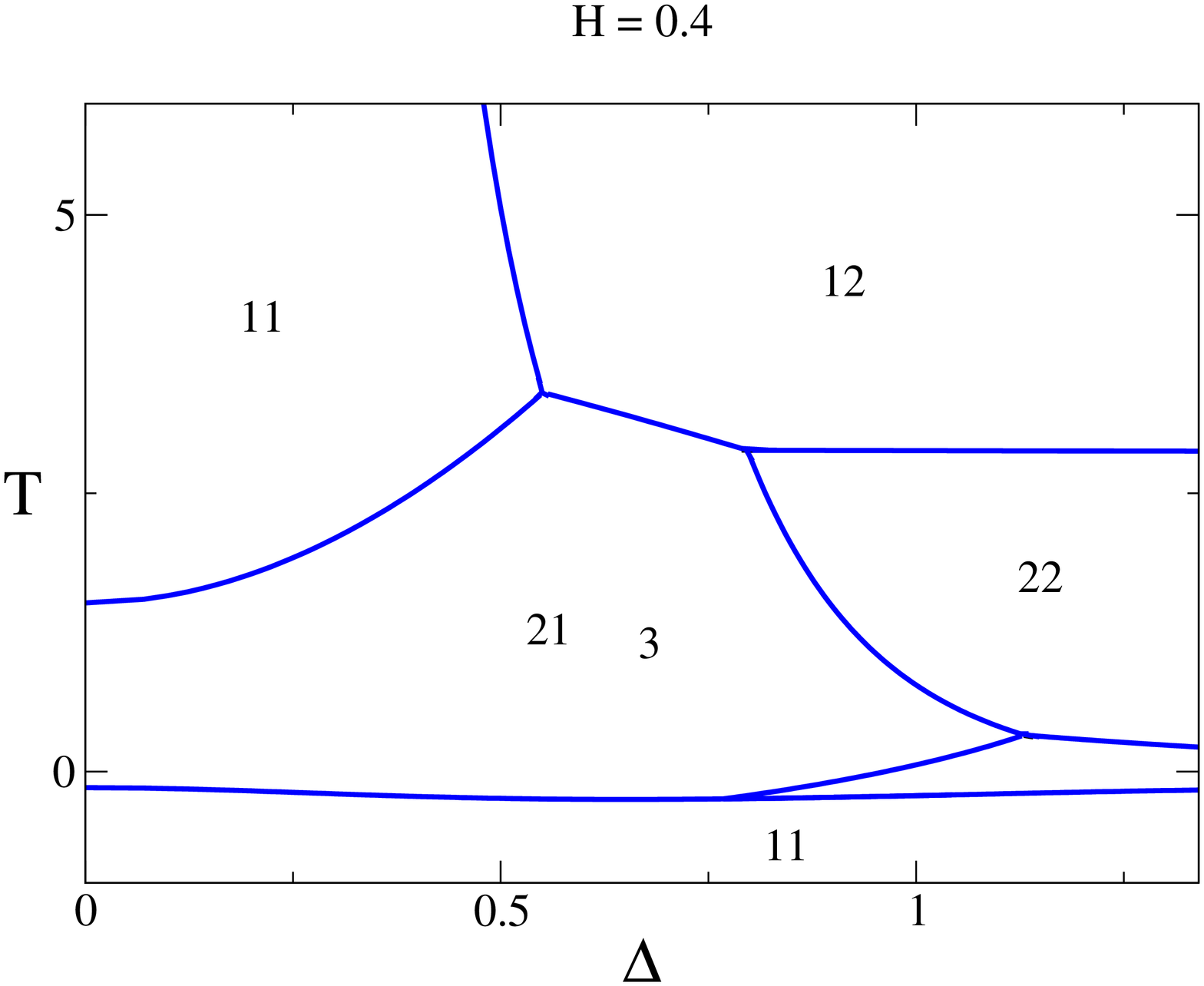}
\caption{Phase diagram for $H = 0.40$.\newline
11 = lamellar parallel,
12 = lamellar perpendicular,\newline
21 = hexagonal parallel,
22 = hexagonal perpendicular,
3  = bcc}
\label{Hp40}
\end{center}
\end{figure}

\begin{figure}[t]
\begin{center}
\includegraphics*[width=90mm,angle=270]{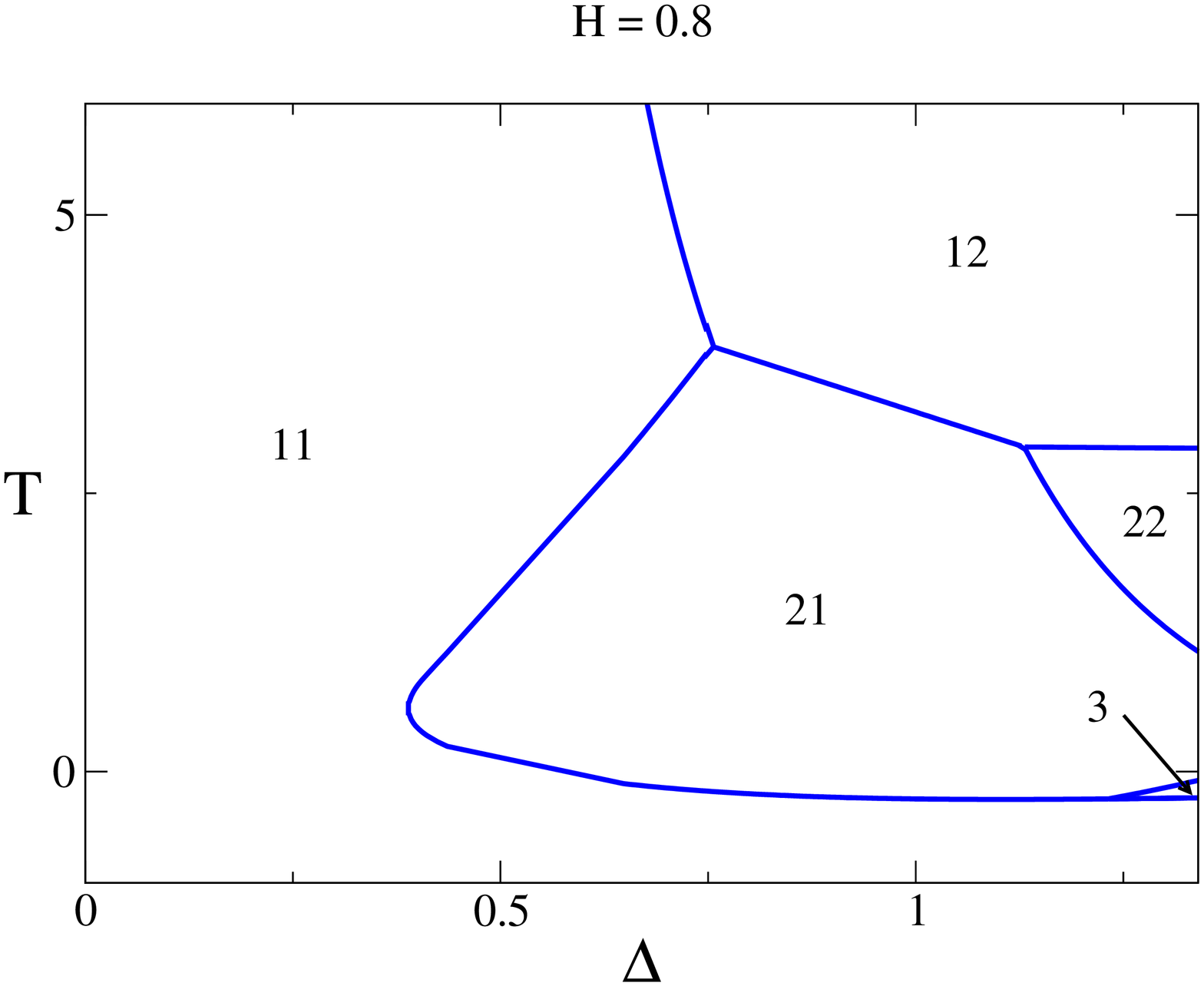}
\caption{Phase diagram for $H = 0.80$.\newline
11 = lamellar parallel,
12 = lamellar perpendicular,\newline
21 = hexagonal parallel,
22 = hexagonal perpendicular,
3  = bcc}
\label{Hp80}
\end{center}
\end{figure}

The results are shown in Figures (\ref{Hm80}) to (\ref{Hp80}). We will start with the simplest case $H=0$ (Figure \ref{Hp0}), where the plates have no preference 
for either monomer type. If the film thickness is an integer, corresponding to $\Delta = 0$, the phase behavior is identical to that of a bulk system,
because the weak segregation bulk equilibrium structures all fit without distortion inside a film of unit thickness. 
We see that on increasing the parameter $T$, which is a measure for the $AB-$repulsion, the
order in which the phases appear is given by homogeneous - bcc - hexagonal - lamellar, which is in accordance with well-known results 
\cite{macromolecules_13_1601_1980}. If the film thickness 
is not an integer, so that $\Delta>0$, the lamellae and the cylinders are oriented perpendicular to the film, and the free energy
of the corresponding structures is independent of the film thickness.
This explains why the phase boundary separating them is a straight horizontal line.
The free energy of the three-dimensional bcc structure, on the other hand, increases if the film thickness deviates from being an integer, because 
it will be compressed or stretched. This is the reason why the bcc structure disappears upon moving to the right in the phase diagram. 

Next consider the phase diagrams for positive values of $H$. We see that close to the line $\Delta = 0$, the lamellae and the cylinders have a
parallel orientation. The reason is that in this way the film can profit optimally from the preference of the plates for the monomers of type $A$
(remember that without affecting the generality we will assume that the $A-$monomers are in the minority). It is for this reason that the border
between the parallel and the perpendicular phases shifts to the right if $H$ is increased. 
Although the perpendicular hexagonal
structure also has a set of perpendicular Fourier vectors which couple to the plates (see Figure (\ref{fourier hexagonal})), 
these vectors couple to the other Fourier vectors only via the (positive) 
fourth order term, while the perpendicular vectors of the parallel hexagonal structure couple favorably to the other Fourier vectors via the (negative) 
third order term. 
The stability of the parallel lamellar structure over the perpendicular lamellar structure for small values of $\Delta$ can also be explained by the positive 
fourth order contribution to the free energy in the latter coming from the coupling between the parallel and the perpendicular Fourier vectors. 
The stronger the plate preference $H$, the more the parallel structures are favored over the perpendicular structures.
More generally, the shifts in the phase boundaries for increasing value of $H$ can be accounted for by differences in the amplitude
of the perpendicular Fourier mode. The structure with the largest amplitude will profit most from an increase in the value of $H$ 
(note that the amplitude may be negative). 

\begin{figure}[t]
\begin{center}
\includegraphics[width=120mm]{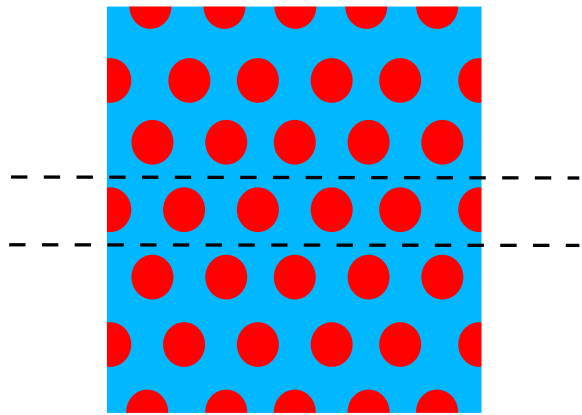}
\caption{Slice through hexagonal phase violating symmetry conditions}
\label{slice_through_hexagonal_structure_violating_symmetries}
\end{center}
\end{figure}

Next we consider the phase diagram for $H<0$. The most striking feature is the total absence of the parallel hexagonal phase. This is due to the fact
that if $H<0$, the third order term and the first order term of the free energy have opposite signs, and compete with each other (see equation \ref{final expressions}). 
In case the minority component forms the cylinders while the majority component forms the matrix, the third order term is negative, but there will be an excess of
the minority component near the plates, which is unfavorable for $H<0$. If, on the other hand, the structure is inverted so that the majority component forms
the cylinders while the minority component forms the matrix, the former will be in excess near the plates, which is favorable, but the price
to be payed is a positive third order contribution. 
Due to this frustration the parallel phase is nowhere stable for $H<0$. One might think that it should be possible 
to cut a slice out of the hexagonal bulk structure such that the cylinders are in the middle of the film, so that the material making up the matrix
is in excess near the plates, preventing the frustration between the first and the third order terms of the free energy expansion. However, it appears
that any slice of integer thickness having the cylinders in the middle violates the symmetry conditions; see, for instance, the failed attempt illustrated 
by Figure (\ref{slice_through_hexagonal_structure_violating_symmetries}). The disappearance of the bcc structure for negative values of $H$ can be accounted 
for by the same mechanism. It should however be noted that for large values of $n$, the alternative way of cutting a slice out of the bcc structure, as illustrated
by the second picture of Figure (), would lead to a lower free energy, so that the corresponding bcc2 structure might remain stable even for negative values of $H$.
Finally we remark that for any point $(\Delta,\chi)$ in the phase diagram, the parallel lamellar structure becomes stable once the absolute value of $H$ 
is large enough. 

Quantitative comparison with experimental work or earlier theoretical work is not well possible, but some of the qualitative aspects of our phase diagrams 
are in accordance with earlier findings. In \cite{Journal_of_Chemical_Physics_118_905_2003} it was remarked that if the surface potential $H$ increases, 
the parallel structures are favored over the perpendicular structures, which is confirmed by our series of phase diagrams. The reason can be found in 
the fact that in the parallel structures, the excess of the minority component at the plates is larger than in the perpendicular ones, while the minority 
component is favored by the plates for positive $H$.  

In \cite{Journal_of_Chemical_Physics_112_2452_2000} it was observed that the lamellar structure gains stability over the hexagonal structure if the surface 
potential increases, which is also apparent from our series of phase diagrams.

In \cite{Journal_of_Chemical_Physics_111_5241_1999} a symmetric diblock copolymer trapped between two parallel plates was investigated using both 
self-consistent field theory, and Monte Carlo
simulations. It was found that in thick films, the lamellae are always oriented parallel to the plates, while for thin films there are transitions
between parallel and perpendicular orientations. Strictly speaking, our phase diagrams are not valid for symmetric copolymers, since we have rescaled
our parameters with powers of $\epsilon$. Of course, the analysis would be easier for symmetric copolymers, and one could represent their phase behavior
by a single diagram in which $\tau / h^{2/3}$ is plotted versus $|D-1|/h^{1/3}$. However, it is possible to understand the occurrence of periodic transitions 
between parallel and perpendicular orientations of the lamellar structure on the basis of our phase diagrams. Suppose that for a given 
copolymer we slowly increase the film thickness, starting from unity. What will be the corresponding path in our phase diagrams? For unit film thickness
we are somewhere on the vertical axis, and we move to the right if the film thickness is increased. However, if we increase the film thickness further, 
at a certain moment the number of periods fitting inside the film will increase jumpwise (either two $1.5$, or to $2$), with as a result that suddenly
the structure is compressed rather than stretched. Increasing the film thickness further wil now lead to a decrease of the parameter $D$ (see the previous section),
so that we start moving to the left in the phase diagram, until we hit the vertical axis once again. It follows that if we keep increasing the film thickness,
we will periodically move to the right and to the left. The turning point at the left will always lie on the vertical axis, but the turning point at the right
will move to the left for thicker films, because the degree of frustration in thick films cannot be very large. Therefore, for thick films the amplitude of
the periodic motion in the phase diagram is so small that the phase boundary between the parallel phase and the perpendicular phase is never reached, so that
the system stays permanently in the parallel phase. For thinner films, however, there might be periodic transitions between the two phases upon increasing the film thickness.
We conclude that our results are in agreement with the results of reference \cite{Journal_of_Chemical_Physics_111_5241_1999}. 

\section{Acknowledgments}
Hindrik Jan Angerman thanks the LEA for financial support.

%\bibliography{references}

\begin{thebibliography}{10}

\bibitem{macromolecules_19_2197_1986}
Thomas, E. {\it et~al.}, {\it Macromolecules}, {\bf 1986}, {\it 19}, 2197.

\bibitem{macromolecules_27_6922_1994}
Foerster, S. {\it et~al.}, {\it Macromolecules}, {\bf 1994} {\it 27}, 6922.

\bibitem{macromolecules_13_1601_1980}
Leibler, L. {\it Macromolecules}, {\bf 1980}, {\it 13}, 1601.

\bibitem{macromolecules_26_6617_1993}
Semenov, A.N. {\it Macromolecules}, {\bf 1993}, {\it 26}, 6617.

\bibitem{Journal_of_Chemical_Physics_106_2436_1997}
Matsen, M.; Bates, F. {\it Journal of Chemical Physics}, {\bf 1997}, {\it 106}, 2436.

\bibitem{Macromol.Symp._81_253_1994}
Erukhimovich, I.; Dobrynin, A. {\it Macromolecular Symposia}, {\bf 1994}, {\it 81}, 253.

\bibitem{PRL_62_1852_1989}
Anastasiadis, S.; Russell, T.; Satija, S.; Majkrzak, C. {\it Physical Review Letters}, {\bf 1989}, {\it 62}, 1852.

\bibitem{macromolecules_33_5702_2000}
Fasolka, M.; Banerjee, P.;Mayes, A. {\it Macromolecules}, {\bf 2000}, {\it 33}, 5702.

\bibitem{PRL_87_035505_2001}
Rehse, N. {\it et~al.}, {\it Physical Review Letters}, {\bf 2001}, {\it 87}, 035505.

\bibitem{macrocolecules_38_4311_2005}
Lee, B. {\it et~al.}, {\it Macromolecules}, {\bf 2005}, {\it 38}, 4311.

\bibitem{PRL_72_2899_1994}
Lambooy, P. {\it et~al.}, {\it Physical Review Letters}, {\bf 1994}, {\it 72}, 2899.

\bibitem{macromolecules_28_7501_1995}
Grim, P.; Nyrkova, I.; G. Semenov, A.N.;  Ten~Brinke, ; G. Hadziioannou,
  {\it Macromolecules}, {\bf 1995}, {\it 28}, 7501.

\bibitem{macromolecules_27_6225_1994}
Walton, D. {\it et~al.}, {\it Macromolecules}, {\bf 1994}, {\it 27}, 6225.

\bibitem{Journal_of_Chemical_Physics_106_7781_1997}
Matsen, M. {\it Journal of Chemical Physics}, {\bf 1997}, {\it 106}, 7781.

\bibitem{macromolecules_30_3097_1997}
Pickett, G. ; Balazs, A. {\it Macromolecules}, {\bf 1997}, {\it 30}, 3097.

\bibitem{Macrocol_Theory_Simul_7_249_1998}
Pickett, G.; Balazs, A. {\it Macrocol. Theory Simul.}, {\bf 1998}, {\it 7}, 249.

\bibitem{Journal_of_Chemical_Physics_111_5241_1999}
Geisinger, T.; Mueller, M.; Binder, K. {\it Journal of Chemical Physics}, {\bf 1999}, {\it 111}, 5241.

\bibitem{European_physical_journal_E_5_605_2001}
Tsori, Y.; Andelman, D. {\it European Physical Journal E}, {\bf 2001}, {\it 5}, 605.

\bibitem{Journal_of_Chemical_Physics_112_2452_2000}
Huinink, H.; Brokken-Zijp, J.C.M.; Van~Dijk, M.A. {\it Journal of Chemical Physics}, {\bf 2000}, {\it 112}, 2452.

\bibitem{Journal_of_Chemical_Physics_118_905_2003}
Szamel, G. {\it Journal of Chemical Physics}, {\bf 2003}, {\it 118}, 905.

\bibitem{physical_review_E_63_061809_2001}
Pareira, G. {\it Physical Review E}, {\bf 2001}, {\it 63}, 061809.

\bibitem{colloid_interface_sciences_90_86_1982}
Silberman, A. {\it Colloid Interface Sci}, {\bf 1982}, {\it 90}, 86.

\bibitem{macromolecules_20_2535_1987}
Fredrickson, G. {\it Macromolecules}, {\bf 1987}, {\it 20}, 2535.

\bibitem{macromolecules_27_4986_1994}
Turner, M.; Rubinstein, M.; Marques, C. {\it Macromolecules}, {\bf 1994}, {\it 27}, 986.

\end{thebibliography}

\clearpage

\newpage

\clearpage

\clearpage

\clearpage

\clearpage

\clearpage

\end{document}